\documentclass{IEEEtran}
\usepackage{amssymb}
\usepackage{amsmath}
\usepackage{graphics}
\usepackage{graphicx}

\input{tcilatex}

\begin{document}
\title{Exact BER Performance Analysis for Downlink NOMA Systems Over
Nakagami-$m$ Fading Channels}
\author{Tasneem Assaf,~Arafat Al-Dweik,~Mohamed El Moursi,~and Hatem
Zeineldin \thanks{%
The authors are with the Department of Electrical and Computer Engineering,
Khalifa University, P.O.Box 127788, Abu Dhabi, UAE. (e-mails:
\{tasneem.assaf, arafat.dweik, mohamed.elmoursi, hatem.zeineldin\}@ku.ac.ae).%
}\thanks{%
A. Al-Dweik is also with Center for Cyber Physical Systems (C2PS), Khalifa
University, Abu Dhabi, UAE, and also with \ the Department of Electrical and
Computer Engineering, Western University, London, ON, Canada. (e-mail:
dweik@fulbrightmail.org).}}
\maketitle

\begin{abstract}
\textbf{In this paper, the performance of a promising technology for the
next generation wireless communications, non- orthogonal multiple access
(NOMA), is investigated. In particular, the bit error rate (BER) performance
of downlink NOMA systems over Nakagami-}$m$\textbf{\ flat fading channels,
is presented. Under various conditions and scenarios, the exact BER of
downlink NOMA systems considering successive interference cancellation (SIC)
is derived. The transmitted signals are randomly generated from quadrature
phase shift keying (QPSK) and two NOMA systems are considered; two users'
and three users' systems. The obtained BER expressions are then used to
evaluate the optimal power allocation for two different objectives,
achieving fairness and minimizing average BER. The two objectives can be
used in a variety of applications such as satellite applications with
constrained transmitted power. Numerical results and Monte Carlo simulations
perfectly match with the derived BER analytical results and provide valuable
insight into the advantages of optimal power allocation which show the full
potential of downlink NOMA systems.}
\end{abstract}

$\boldsymbol{Index}$ $\boldsymbol{Terms:}$ NOMA, BER, SIC, Optimal Power
Allocation, Fairness, Minimum Average BER, Nakagami-$m$.

\section{Introduction}

\markboth{IEEE Journals, Vol. XX, No. Y, 04 Aug. 2019}
{Murray and Balemi: Using the Document Class IEEEtran.cls}%
\setcounter{page}{1}The expeditious development of the mobile Internet and
Internet of Things (IoT) has obtruded several challenges on the Fifth
Generation (5G) and Beyond 5G (B5G) wireless communications systems.\ Such
challenges include capacity increase by a factor of $1000$, data rates
exceeding $10$ Gb/s, $10$ years battery life for machine-to-machine (M2M)
communications, and network latency less than $1$ ms \cite{5G-requirements}.
Consequently, extensive research has been focused in the last few years to
develop the enabling technologies that can satisfy such requirements, which
include dense heterogeneous networks, full-duplex communication,
energy-aware communication and energy harvesting 
\cite{Lina-SWIPT, Ekram-5G-enabling}%
, cloud-based radio access networks \cite{Gharaybeh-IEEE-Access}, wireless
network virtualization \cite{Gharaybeh-IEEE-Lett}, advanced antenna systems 
\cite{Alouini-massive-mimo}, and efficient error correction coding \cite%
{Arafat-Survey}. Moreover, great advancements have been achieved in terms of
multiple access technologies such as the non-orthogonal multiple access
(NOMA), which may improve the spectral efficiency and latency with respect
to orthogonal multiple access (OMA) \cite{8357810}. Several NOMA schemes
have been actively investigated, which can be categorized into two main
categories, power-domain NOMA, \cite{7676258}-\cite{8588360} and code-domain
NOMA \cite{5956554}-\cite{8515248}. The focus of this paper is on the
power-domain NOMA.

The performance of NOMA systems has attracted extensive literature, which is
mainly focus outage probability and capacity. For example, the outage
achievable rate region is studied in \cite{8466830}, where the results imply
that NOMA outperforms OMA under similar conditions. In \cite{7095538}, the
authors investigate a two-users NOMA\ system in terms of their power
allocation, through the maximization of the ergodic sum capacity with the
constraint of minimum sum rate requirement, fixed total transmit power, and
partial channel state information (CSI) availability. In \cite{7564993}, a
novel NOMA clustering scheme using a power allocation mechanism is presented
to reduce the computational complexity at the expense of user fairness
compared to the full search method. In \cite{7510916}, with the objective of
providing proportional fairness, optimal power allocation and user pairing
problems in multiuser downlink NOMA are investigated. A sum rate comparison
between multiple-input multiple-output (MIMO) NOMA and OMA clusters is
conducted in \cite{7236924}, where each cluster consists of two users. In 
\cite{5672457}, various user scheduling strategies are presented to
accomplish flexible and efficient trade-offs between capacity and user
fairness in NOMA systems. In \cite{7273963} and \cite{7838583}, user
clustering and power allocation for downlink NOMA system have been studied
thoroughly.

More recently, noticeable efforts have focused on evaluating the bit error
rate (BER) performance of NOMA systems with imperfect successive
interference cancellation (SIC). For example, the authors of \cite{8501953}
derived closed-form expressions for the union bound on the BER of downlink
NOMA with imperfect SIC over Nakagami-$m$ fading channels. Although the
derived bounds are useful to estimate the BER, the results presented in \cite%
{8501953} show that the bounds may deviate significantly from the simulation
results in certain scenarios. The average BER performance of a NOMA system
using space-shift keying (SSK) in Rayleigh fading channels is investigated
in \cite{8703784} where the exact BER is expressed in closed-form only for
users two and three in a three users scenario. The exact symbol error rate
(SER) for a downlink\ NOMA with imperfect SIC is presented in \cite%
{SER-TVT-2019}. Nevertheless, using the BER is more informative when
comparing different systems with different modulation orders, and the
results are limited only to the two users scenario and Rayleigh fading
channels. The exact SER for the two users scenario in Rayleigh fading
channels is also investigated in \cite{8733878}. The BER of uplink NOMA for
the two users scenario is considered in \cite{7967766} under imperfect SIC
scenarios. The main limitation of this work is that the channel fading is
considered constant over the transmission block, and hence, the channel
becomes effectively an additive white Gaussian noise (AWGN) channel. An
asynchronous uplink NOMA system based on triangle-SIC\ error is presented in 
\cite{7817823}. Similar to \cite{7967766}, the channel coefficients are
assumed to be fixed, hence the derived closed-form expressions can not be
used in random fading scenarios. In \cite{8457925}, the BER is derived for a
two users downlink\ and uplink NOMA systems over Rayleigh fading channels.
However, the assumption that the links in downlink NOMA follows independent
and identically distributed (i.i.d.) Rayleigh fading overlooks the large
scale fading factor, which limits the contribution of this work. Moreover,
the paper lacks in-depth insights into the analysis of the obtained BER\
results.

\subsection{Motivation and Main contributions}

As can be noted from the aforementioned literature survey, to the best of
the authors' knowledge, there is no work reported that considers the exact
BER analysis of downlink NOMA in Nakagami-$m$ fading channels. Moreover, the
works considered the special case where $m=1$, i.e. Rayleigh, focus on the
two-users scenario. Therefore, the main contributions of this paper are
summarized as follows:

\begin{enumerate}
\item The exact BER performance analysis of a downlink NOMA with imperfect
SIC over Nakagami-$m$ flat fading channel is considered, where exact
analytical BER expressions are derived for each user individually for the
cases of two and three users' scenarios. The derived BER expressions are
verified by Monte Carlo simulations.

\item The exact BER is derived in terms of a closed-form expressions for the
special case of $m=1$, Rayleigh fading, for two and three users scenarios.

\item The optimal power allocation for all users is investigated based on
the derived BER expressions for two different criteria. In the first, the
power allocation for each user is allocated optimally to guarantee fairness
among all users, which is expressed in terms of equal BER for all users. In
the second, the power allocation coefficients are selected optimally to
minimize the average BER for all users.
\end{enumerate}

\subsection{Notations}

To notations used throughout the paper are as follows. $\Pr(\cdot )=\mathrm{%
P\!}\left( \cdot\right) $, $\mathrm{P}\left( a,b\right) =\mathrm{P\!}\left(
a\cap b\right) $, $\mathrm{P\!}\left( a;b\right) =\mathrm{P\!}\left( a\cup
b\right) $, $\mathrm{P\!}\left( s_{1}=a_{c},s_{2}=a_{k},s_{3}=a_{v}\right) $ 
$\rightarrow$ $\mathrm{P\!}\left( a_{c},a_{k},a_{v}\right) $, $%
P_{b_{ni}}\triangleq\mathrm{P\!}\left( \hat {b}_{ni}\neq b_{ni}\right) $, $%
b_{ni}=c\rightarrow b_{ni}^{(c)}$, $c\in\left\{ 0,1\right\} $, and $%
s_{n}=a_{i}\rightarrow s_{n}^{(i)}$, $i\in\left\{ 0,1,2,3\right\} $.

\subsection{Paper organization}

The rest of the paper is organized as follows. In Sec. \ref%
{SystemandCHannelModel}, the system and channel models are presented. This
is followed by exact BER analysis for the two-users and three-users downlink
NOMA\ systems are presented in Sec. \ref{Sec-exact BER} and Sec. \ref%
{SubSec-BER U1-3U}, respectively. The optimal power allocation problem is
formulated in Sec. \ref{Optimality}, while numerical and simulation results
are shown in Sec. \ref{Results}. Finally, the work is concluded in Sec. \ref%
{Conclusion}.

\section{\label{SystemandCHannelModel}System and Channel Models}

This work considers a power-domain downlink NOMA system with $N$ users, $%
U_{1}$, $U_{2}$, $\ldots$, $U_{N}$. The users' equipment (UEs) and the base
station (BS) are equipped with single antennas \cite{8501953}. Therefore,
the transmitted signal from the BS can be expressed as%
\begin{equation}
x=\dsum \limits_{n=1}^{N}\sqrt{\beta_{n}P_{T}}s_{n}  \label{MLD_U1}
\end{equation}
where $s_{n}$ is the information signal of the $n$th user selected uniformly
from a particular symbol constellation, $P_{T}$ is the BS transmit power,
and $\beta_{n}$ is the allocated power coefficient for the $n$th user such
that $\dsum _{n=1}^{N}\beta_{n}=1$. For the rest of the paper, the transmit
power $P_{T}$ is normalized to unity.

In flat fading channels, the received signal at the $n$th UE can be written
as 
\begin{equation}
r_{n}=h_{n}x+w_{n}
\end{equation}
where $h_{n}$ represents the link between the BS and the $n$th user whose
probability density function (PDF) is described in Appendix I and $w_{n}$ is
the AWGN, $w_{n}\sim\mathcal{CN}\left( 0,N_{0}\right) .$ Given that the
channel phase $\theta_{n}$ is estimated and compensated perfectly at the
receiver, then the received signal after phase compensation $\check{r}%
_{n}=r_{n}\mathrm{e}^{-j\theta_{n}}=\alpha_{n}x+\check{w}_{n}$, where $%
\check{w}_{n}=w_{n}\mathrm{e}^{-j\theta_{n}}$ and $\alpha_{n}=\left\vert
h_{n}\right\vert $ is the channel gain. Assuming that the AWGN\ is
circularly symmetric, then $\check{w}_{n}$ and $w_{n}$ have identical PDFs,
consequently $\check{w}_{n}$ and $w_{n}$ can be used interchangeably.
Without loss of generality, it is assumed that the first user has the lowest
channel gain, and the second user has the second lowest channel gain, and so
forth, i.e., $\alpha_{1}<\alpha_{2}<\cdots<\alpha_{N}$. To enable reliable
detection of all users, it is necessary to cancel the inter-user
interference (IUI), which is typically performed using SIC. Therefore, the
power should be allocated in the opposite order of the channel gains, i.e., $%
\beta_{1}>\beta_{2}>\cdots >\beta_{N}$.

To detect the signal of the $n$th user, the signals of $U_{1}$, $U_{2}$, $%
\ldots$, $U_{n-1}$ should be detected and scaled, then subtracted from $%
r_{n} $, the IUI for users $U_{n+1}$, $U_{n+2}$, $\ldots$, $U_{N}$ is
considered as unknown additive noise. For the first user, the IUI from all
users will be treated as noise, and thus, the maximum likelihood detector
(MLD) given that the channel gain $h_{1}$ is known perfectly at the receiver
can be expressed as \cite{Proakis2007},%
\begin{equation}
\hat{s}_{1}=\arg\min_{\tilde{s}_{1}\in\mathbb{S}}\left\vert r_{1}-\sqrt {%
\beta_{1}}h_{1}\tilde{s}_{1}\right\vert ^{2}  \label{firstuser_MLD}
\end{equation}
where $\hat{s}_{1}$ is the estimated data symbol, $\mathbb{S}$ is the set of
all possible constellation points for $U_{1}$, and $\tilde{s}_{1}$ are the
trial values of $s_{1}$. For the $n$th user, the detector can be described by%
\begin{equation}
\hat{s}_{n}=\arg\min_{\tilde{s}_{n}\in\mathbb{S}}\left\vert \left(
r_{n}-h_{n}\dsum \limits_{k=1}^{n-1}\sqrt{\beta_{k}}\hat{s}_{k}\right) -%
\sqrt{\beta_{n}}h_{n}\tilde{s}_{n}\right\vert ^{2}.  \label{MLD_general}
\end{equation}

In the following sections, the exact BER is derived for a power-domain NOMA
system with two and three users. Although the presented approach can be
applied to any phase shift keying (PSK) or quadrature amplitude modulation
(QAM), the derivation becomes intractable for a modulation order $M>4$,
particularly for $N>2$. Therefore, the analysis presented in this work
considers Gray coded quadrature PSK (QPSK) modulation where $\mathbb{S=}%
\left\{ a_{0}=00\text{, }a_{1}=01\text{, }a_{2}=10\text{, }a_{3}=11\right\} $%
, and all symbols are considered equiprobable.

\section{\label{Sec-exact BER}NOMA Bit Error Rate (BER) Analysis: Two Users (%
$N=2)$}

\begin{figure}
    \centering
    \includegraphics{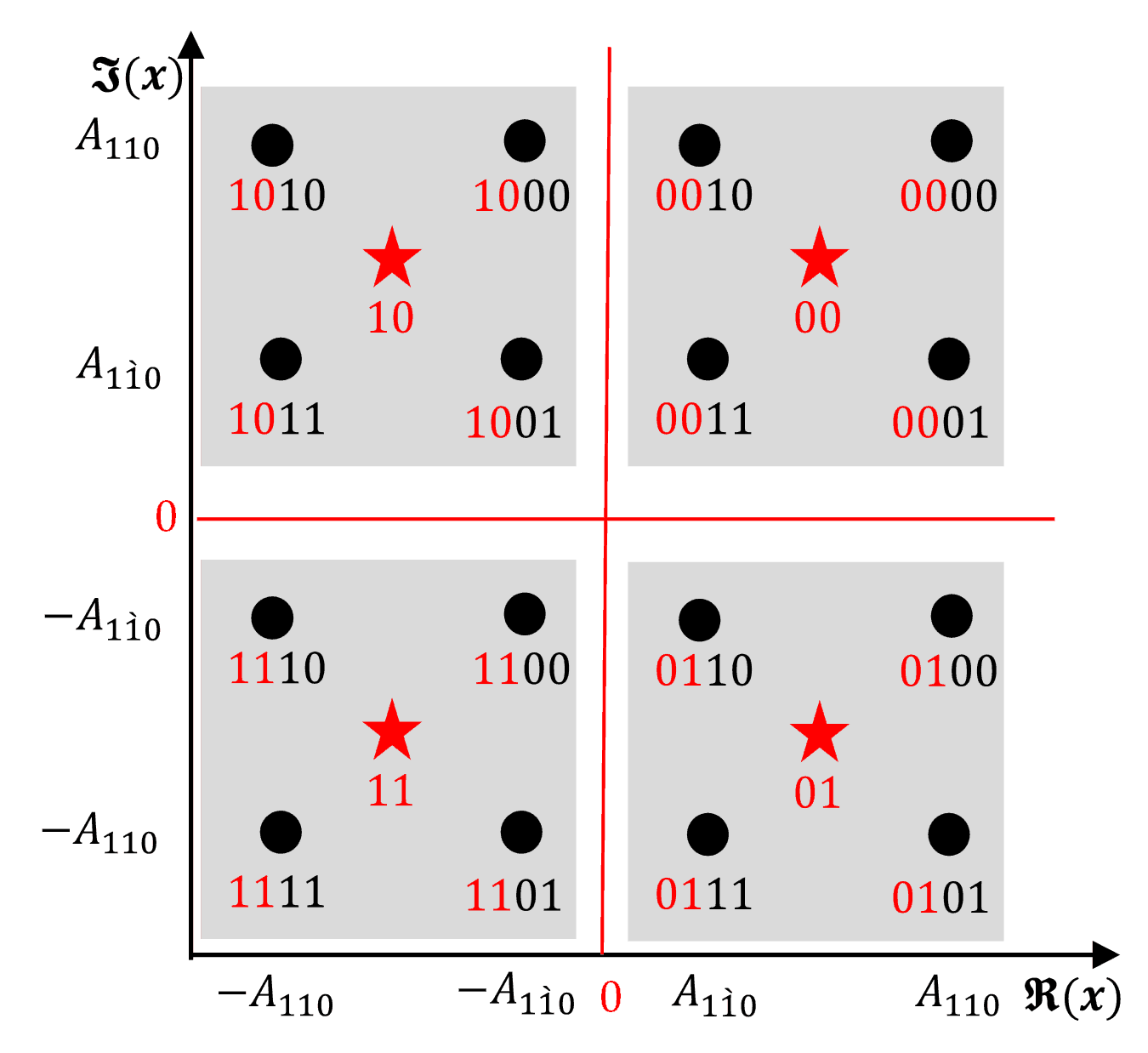}
    \caption{The constellation diagram of the transmitted symbol $x$ for $N=2$.}
    \label{general2u}
\end{figure}

As can be noted from (\ref{MLD_U1}), the transmitted symbol $x$ for $N=2$ is
the superposition of two QPSK symbols, and hence, it should correspond to
one of the $16$ constellation points shown in Fig. \ref{general2u}. The bit
representation for each constellation point is given in the form of $%
\begin{bmatrix}
b_{11} & b_{12} & b_{21} & b_{22}%
\end{bmatrix}
$, for each bit $b_{ni}$, $\left\{ n,i\right\} \in\left\{ 1\text{, }%
2\right\} $, where $n$ denotes the user index while $i$ denotes the bit
index.


\subsection{\label{SubSec-BER U1}BER of the first user $\left(
U_{1}|_{N=2}\right) $}

For the first user, the detection is performed using (\ref{firstuser_MLD}),
and thus, no SIC is required. Based on the specific value of $s_{2}$, the
IUI caused by $U_{2}$, the symbol $x$ may become one of the four
constellation points in the neighborhood of $s_{1}$. The shaded blocks in
Fig. \ref{general2u} show the four possible values that a particular symbol $%
s_{1}$ may take. For example, given that $s_{1}=10$ then $%
x|_{s_{1}}\in\left\{ 1000\text{, }1001\text{, }1010\text{, }1011\right\} $.
The amplitudes of the inphase $x_{I}\triangleq\Re\left( x\right) $ and
quadrature $x_{Q}\triangleq\Im\left( x\right) $ for each constellation point
are defined as 
\begin{equation}
A_{u_{1}u_{2}u_{3}}=u_{1}\sqrt{\beta_{1}}+u_{2}\sqrt{\beta_{2}}+u_{3}\sqrt{%
\beta_{3}}\text{, }u_{i}\in\left\{ 0,1,\acute{1},2\right\}
\end{equation}
where $\acute{1}\triangleq-1$. For example, given that $%
s_{1}^{(2)}=s_{2}^{(2)}$, then $x=1010$ and $x_{I}=-\sqrt{\beta_{1}}-\sqrt{%
\beta_{2}}\triangleq A_{\acute{1}\acute{1}0}$, and $x_{Q}=\sqrt{\beta_{1}}+%
\sqrt {\beta_{2}}\triangleq A_{110}$. It is worth noting that for $N=2$, $%
u_{3}=0$ regardless the values of $s_{1}$ or $s_{2}$, however, it is used to
unify the notation throughout the paper.

The probability of error for each bit actually depends on the values of $%
s_{1}$ and $s_{2}$. For example, given that $s_{1}^{(2)}$ , the first bit $%
b_{11}$ might be detected incorrectly if $\hat{s}_{1}=a_{0}$ $(00)$ or $%
a_{1} $ $(01)$, as shown in Fig. \ref{general2u}. However, $\mathrm{P\!}%
\left( \hat{s}_{1}=a_{0}\text{ or }a_{1}\right) $, denoted as $\mathrm{P\!}%
\left( \hat{s}_{1}=a_{0};a_{1}\right) ,$ depends on $s_{2}$ as well.
Therefore, the average BER should consider all possible combinations of $%
s_{1}$ and $s_{2}$,%
\begin{equation}
P_{b_{1i}}=\sum_{l,k}\left( P_{b_{1i}}|_{s_{1}^{(l)}\text{, }%
s_{2}^{(k)}}\right) \mathrm{P\!}\left( s_{1}^{(l)}\text{, }%
s_{2}^{(k)}\right) .  \label{E-Pr-U1bi}
\end{equation}
By noting that $s_{1}$ and $s_{2}$ are independent, then (\ref{E-Pr-U1bi})
can be written as,%
\begin{equation}
P_{b_{1i}}=\frac{1}{16}\sum_{\{l.k\}=0}^{3}\left( P_{b_{1i}}|_{s_{1}^{(l)}%
\text{, }s_{2}^{(k)}}\right) .  \label{E_Pb_U1_bi}
\end{equation}

$\boldsymbol{Case}$ $\boldsymbol{1}$: $s_{1}^{(2)}$, $s_{2}^{(0)}$: For this
case, $x|_{a_{2},a_{0}}=-A_{1\acute{1}0}+jA_{110}$, to simplify th notations 
$x|_{a_{l},a_{k}}$ is written as $x_{l,k}$. Consequently, the error
probability of $b_{11}$ is given by, 
\begin{equation}
P_{b_{11}}|_{s_{1}^{(2)}\text{, }s_{2}^{(0)}}=\mathrm{P\!}\left( \hat{s}%
_{1}=a_{0};a_{1}\right) .  \notag
\end{equation}
As can be noted from Fig. \ref{general2u}, $\mathrm{P\!}\left( \hat{s}%
_{1}=a_{0};a_{1}\right) $ depends only on the inphase component of $\check {r%
}_{1}$, i.e., $\Re\left( \check{r}_{1}\right) \triangleq\mathfrak{r}_{1}$
and the specific value of $x$. Thus,%
\begin{align}
P_{b_{11}}|_{s_{1}^{(2)}\text{, }s_{2}^{(0)}} & =\mathrm{P\!}\left( 
\mathfrak{r}_{1}\geq0\right)  \notag \\
& =\mathrm{P\!}\left( -\alpha_{1}A_{1\acute{1}0}+\mathfrak{n}_{1}\geq0\right)
\notag \\
& =\mathrm{P\!}\left( \mathfrak{n}_{1}\geq\alpha_{1}A_{1\acute{1}0}\right)
\label{(E1)}
\end{align}
where $\mathfrak{r}_{1}=-\alpha_{1}A_{1\acute{1}0}+\mathfrak{n}_{1}$, $%
\Re\left( \check{w}_{1}\right) \triangleq\mathfrak{n}_{1}$. Therefore, 
\begin{align}
P_{b_{11}}|_{s_{1}^{(2)}\text{, }s_{2}^{(0)}} & =\frac{1}{\sqrt{2\pi \sigma_{%
\mathfrak{n}_{1}}^{2}}}\int_{\alpha_{1}A_{1\acute{1}0}}^{\infty }e^{-\frac{%
z^{2}}{2\sigma_{\mathfrak{n}_{1}}^{2}}}d\mathfrak{n}_{1}  \notag \\
& =Q\left( \sqrt{\gamma_{1,1}}\right)  \label{E2}
\end{align}
where $\gamma_{1,1}=\alpha_{1}^{2}A_{1\acute{1}0}^{2}/\sigma_{\mathfrak{n}%
_{1}}^{2}$ and $Q\left( .\right) $ denotes the Gaussian $Q$ function.

$\boldsymbol{Case}$ $\boldsymbol{2}$: $s_{1}^{(2)}$, $s_{2}^{(1)}$:

This case is similar to the case of $s_{2}^{(0)}$, hence, the error
probability is given by (\ref{E2}) as well.

$\boldsymbol{Case}$ $\boldsymbol{3}$: $s_{1}^{(2)}$, $s_{2}^{(2)}$:

In this case, $x_{2,2}=-A_{110}+jA_{110}$, then the error probability can be
expressed as%
\begin{align}
P_{b_{11}}|_{s_{1}^{(2)}\text{, }s_{2}^{(2)}} & =\mathrm{P\!}\left( 
\mathfrak{r}_{1}\geq0\right)  \notag \\
& =\mathrm{P\!}\left( \mathfrak{n}_{1}\geq\alpha_{1}A_{110}\right) .
\end{align}
Following the same approach used to derive (\ref{E2}) gives, 
\begin{equation}
P_{b_{11}}|_{s_{1}^{(2)}\text{, }s_{2}^{(2)}}=Q\left( \sqrt{\gamma_{1,2}}%
\right)
\end{equation}
where $\gamma_{1,2}=\alpha_{n}^{2}A_{110}^{2}/\sigma_{\mathfrak{n}_{1}}^{2}$.

$\boldsymbol{Case}$ $\boldsymbol{4}$: $s_{1}^{(2)}$, $s_{2}^{(3)}$:

The probability of error in this case is similar to the case of $s_{1}^{(2)}$%
, $s_{2}^{(2)}$.

The remaining cases, $\boldsymbol{Case}$ $\boldsymbol{5}$ to $\boldsymbol{%
Case}$ $\boldsymbol{16}$ are similar to $\boldsymbol{Case}$ $\boldsymbol{1}$
to $\boldsymbol{Case}$ $\boldsymbol{4}$ except that the value of $s_{1}$ is
replaced by $a_{0}$, $a_{1}$, $a_{2}$ and $a_{3}$. Substituting the results
of the $16$ cases in (\ref{E_Pb_U1_bi}) gives $P_{11}=\frac{1}{2}\left[
Q\left( \sqrt{\gamma_{1,1}}\right) +Q\left( \sqrt{\gamma_{1,2}}\right) %
\right] $. It is also straightforward to show that $P_{12}=P_{11}$.
Therefore, the conditional BER of the first user is given:%
\begin{align}
P_{U_{1}} & =\frac{1}{2}\left[ P_{b_{11}}+Pb_{_{12}}\right]  \notag \\
& =\frac{1}{2}\left[ Q\left( \sqrt{\gamma_{1,1}}\right) +Q\left( \sqrt{%
\gamma_{1,2}}\right) \right] .  \label{E3}
\end{align}

\subsection{\label{SubSec-BER U2}BER of Second User $\left(
U_{2}|_{N=2}\right) $}

\begin{figure}
    \centering
    \includegraphics{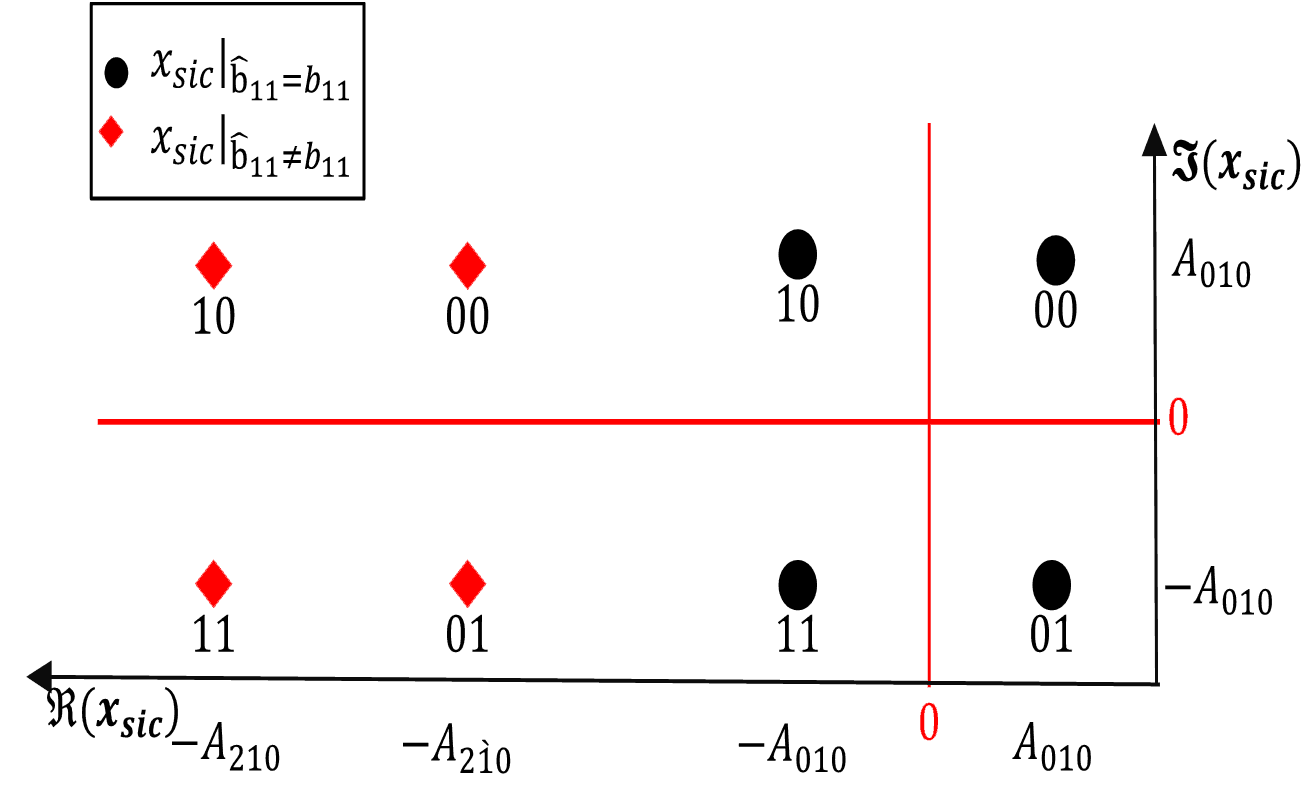}
    \caption{Equivelant constellation of $x_{sic}|_{\hat{s%
}_{1}=s_{1}}$ and $x_{sic}|_{\hat{s}_{1}\neq s_{1}}$ of the second user, $N=2
$.}
    \label{2U_cases_U2}
\end{figure}

To detect its own symbol $s_{2}$, the second user should initially detect $%
s_{1}$ as described in (\ref{firstuser_MLD}), and then compute,%
\begin{align}
\hat{s}_{2} & =\arg\min_{\tilde{s}_{2}\in\mathbb{S}}\left\vert r_{2,sic}-%
\sqrt{\beta_{2}}h_{2}\tilde{s}_{2}\right\vert ^{2}  \notag \\
& =\arg\min_{\tilde{s}_{2}\in\mathbb{S}}\left\vert x_{sic}h_{2}+w_{2}-\sqrt{%
\beta_{2}}h_{2}\tilde{s}_{2}\right\vert ^{2}  \label{E-13}
\end{align}
where $r_{2,sic}=r_{2}-h_{2}\sqrt{\beta_{1}}\hat{s}_{1}$ and $x_{sic}=x-%
\sqrt{\beta_{1}}\hat{s}_{1}$. Therefore, given that $\hat{s}_{1}=s_{1}$,
then $x_{sic}=\sqrt{\beta_{2}}s_{2}$ and $r_{2,sic}$ corresponds to IUI-free
QPSK signal. The constellation diagram of $x_{sic}|_{\hat{s}_{1}=s_{1}}$ is
shown in Fig. \ref{2U_cases_U2}. On the other hand, if $\hat{s}_{1}\neq
s_{1} $, then $x_{sic}=\sqrt{\beta_{1}}s_{1}+\sqrt{\beta_{2}}s_{2}-\sqrt{%
\beta_{1}}\hat{s}_{1}$ and its constellation diagram depends on $\hat{s}_{1}$%
. For example, given that $s_{1}^{(0)}$, $\hat{s}_{1}^{(2)}$, the
constellation diagram of $x_{sic}$ becomes as shown in Fig. \ref{2U_cases_U2}%
.

The BER of $U_{2}$ depends on $s_{1}$, $s_{2}$ and $\hat{s}_{1}$.
Therefore, the probability of error should be averaged over all possible
combinations,%
\begin{equation}
P_{b_{2i}}=\sum_{v,k,l}P_{b_{2i}}|_{s_{1}^{(v)},s_{2}^{(k)},\hat{s}%
_{1}^{(l)}}\mathrm{P\!}\left( s_{1}^{(v)},s_{2}^{(k)},\hat{s}%
_{1}^{(l)}\right) .  \label{E-Pr-U2bi}
\end{equation}
Using the chain rule $\mathrm{P\!}\left( s_{1}^{(v)},s_{2}^{(k)},\hat{s}%
_{1}^{(l)}\right) =\mathrm{P\!}\left( \hat{s}%
_{1}^{(a_{l})}|_{s_{1}^{(v)},s_{2}^{(k)}}\right) \mathrm{P\!}\left(
s_{1}^{(v)},s_{2}^{(k)}\right) $ and noting that $s_{1}$ and $s_{2}$ are
independent, then (\ref{E-Pr-U2bi}) can be written as, 
\begin{equation}
P_{b_{2i}}=\frac{1}{16}\sum_{\left\{ k,l,v\right\}
=0}^{3}P_{b_{2i}}|_{s_{1}^{(v)},s_{2}^{(k)},\hat{s}_{1}^{(l)}}\mathrm{P\!}%
\left( \hat{s}_{1}^{(a_{l})}|_{s_{1}^{(v)},s_{2}^{(k)}}\right) .
\label{Eb21}
\end{equation}
It should be noted that the probability of correct or incorrect detection of 
$b_{12}$ does not affect the error probability of the second user bits $%
P_{2i}$. Therefore, it is assumed that $P_{b_{12}}=1$ for all cases.

The case where $v=l$ corresponds to the event that $\hat{s}_{1}=s_{1}$, and
the corresponding probabilities can be computed as follows:

$\boldsymbol{Case}$ $\boldsymbol{1}$: $s_{1}^{(0)}\left( b_{11}^{(0)}\right) 
$, $\hat{s}_{1}^{(0)}\left( \hat{b}_{11}^{(0)}\right) $, $s_{2}^{(2)}$:

The probability $\mathrm{P\!}\left( \hat{b}_{11}^{(0)}=0|_{b_{11}^{(0)}\text{%
,}s_{2}^{(2)}}\right) $ can be obtained by considering the error probability
of $U_{1}$ given that $b_{11}^{(0)}=\hat{b}_{11}^{(0)}$ $\left(
s_{1}^{(0)}\right) $ and $s_{2}^{(2)}$. In this case, the conditional error
probability of $b_{21}$ can be computed by noting that $P_{b_{21}}|\mathbb{A}%
_{221}=\mathrm{P\!}\left( \hat{s}_{2}=a_{0};a_{1}|_{\mathbb{A}_{221}}\right) 
$ where $\mathbb{A}_{221}\rightarrow\left\{ b_{11}^{(0)},\hat{b}%
_{11}^{(0)},s_{2}^{(2)}\right\} $. The transmitted signal amplitude for this
case is $x=A_{1\acute{1}0}+jA_{110}$, thus, $x_{sic}=A_{0\acute{1}%
0}+jA_{010} $ and $r_{2,sic}=\alpha_{2}x_{sic}+w_{2}.$ As can be noted from
Fig. \ref{2U_cases_U2}, $\mathrm{P\!}\left( \hat{s}_{2}=a_{0};a_{1}|_{%
\mathbb{A}_{221}}\right) $ depends only on the inphase component of $\check{r%
}_{2,sic}$, i.e., $\Re\left( \check{r}_{2,sic}\right) \triangleq\mathfrak{r}%
_{2,sic}$. Thus 
\begin{align}
P_{b_{21}}|_{\mathbb{A}_{221}} & =\mathrm{P\!}\left( \mathfrak{n}%
_{2}+\alpha_{2}A_{0\acute{1}0}\geq0\left\vert \mathfrak{n}_{2}+\alpha
_{2}A_{1\acute{1}0}\geq0\right. \right)  \notag \\
& =\mathrm{P\!}\left( \mathfrak{n}_{2}\geq\alpha_{2}A_{010}\left\vert 
\mathfrak{n}_{2}\geq\alpha_{2}A_{\acute{1}10}\right. \right)  \label{Eb1111}
\end{align}
where $\mathfrak{r}_{2,sic}=\alpha_{2}A_{0\acute{1}0}+\mathfrak{n}_{2}$, $%
\Re\left( \check{w}_{2}\right) \triangleq$ $\mathfrak{n}_{2}$and $-A_{0%
\acute{1}0}=A_{010}$. Using Bayes' theorem and considering the results in (%
\ref{Eb21}) and (\ref{Eb1111}), we obtain%
\begin{align}
P_{b_{21}}|\mathbb{A}_{221}\mathrm{P\!}\left( \hat{b}_{11}^{\left( 0\right)
}|_{b_{11}^{(0)},s_{2}^{(2)}}\right) & =\mathrm{P\!}\left( \mathfrak{n}%
_{2}\geq\alpha_{2}A_{\acute{1}10}\right)  \notag \\
& \times\mathrm{P\!}\left( \mathfrak{n}_{2}\geq\alpha_{2}A_{010}\left\vert 
\mathfrak{n}_{2}\geq\alpha_{2}A_{\acute{1}10}\right. \right)
\label{E-P21-01}
\end{align}
By noting that the right hand side of (RHS) of (\ref{E-P21-01}) is equal to $%
\mathrm{P\!}\left( \mathfrak{n}_{2}\geq\alpha_{2}A_{010}\text{, }\mathfrak{n}%
_{2}\geq\alpha_{2}A_{\acute{1}10}\right) $, then 
\begin{align}
P_{b_{21}}|\mathbb{A}_{221}\mathrm{P\!}\left( \hat{b}%
_{11}^{(0)}|_{b_{11}^{(0)}\text{,}s_{2}^{(2)}}\right) & =\mathrm{P\!}\left( 
\mathfrak{n}_{2}\geq\alpha_{2}A_{010}\right)  \notag \\
& =Q\left( \sqrt{\gamma_{2,1}}\right)
\end{align}
where $\gamma_{2,1}=\alpha_{2}^{2}A_{010}^{2}/\sigma_{\mathfrak{n}_{2}}^{2}$.

For $b_{22}$, the error probability is obtained using the approach used with 
$b_{21}$ 
\begin{multline}
P_{b_{22}}|\mathbb{A}_{221}\mathrm{P\!}\left( \hat{b}%
_{11}^{(0)}|_{b_{11}^{(0)}\text{,}s_{2}^{(2)}}\right) =\mathrm{P\!}\left( 
\mathfrak{n}_{2}\geq\alpha_{2}A_{\acute{1}10}\right)  \label{E-Pb22-00} \\
\times\mathrm{P\!}\left( \mathfrak{q}_{2}+\alpha_{2}A_{010}\leq0\left\vert 
\mathfrak{n}_{2}\geq\alpha_{2}A_{\acute{1}10}\right. \right) .
\end{multline}
As the RHS of (\ref{E-Pb22-00}) can be simplified to $\mathrm{P\!}\left( 
\mathfrak{q}_{2}\leq\alpha_{2}A_{0\acute{1}0},\mathfrak{n}_{2}\geq\alpha
_{2}A_{\acute{1}10}\right) $, then 
\begin{equation}
P_{b_{22}}|\mathbb{A}_{221}\mathrm{P\!}\left( \hat{b}%
_{11}^{(0)}|_{b_{11}^{(0)},s_{2}^{(2)}}\right) =Q\left( \sqrt{\gamma_{2,1}}%
\right) \left( 1-Q\left( \sqrt{\gamma_{2,2}}\right) \right)
\end{equation}
where $\mathfrak{q}_{2}\triangleq\mathfrak{I}\left( \check{w}_{2}\right) $
and $\gamma_{2,2}=\alpha_{2}^{2}A_{1\acute{1}0}^{2}/\sigma_{\mathfrak{n}%
_{2}}^{2}.$

$\boldsymbol{Case}$ $\boldsymbol{2}$: $s_{1}^{(0)}\left( b_{11}^{(0)}\right) 
$, $\hat{s}_{1}^{(0)}$ $\left( \hat{b}_{11}^{(0)}\right) $, $s_{2}^{(0)}$:

In this case, $\left( P_{21}|_{\mathbb{A}_{222}}\right) \mathrm{P\!}\left( 
\hat{b}_{11}^{(0)}|_{b_{11}^{(0)},s_{2}^{(0)}}\right) $ where $\mathbb{A}%
_{222}\rightarrow\left\{ b_{11}^{(0)},\hat{b}_{11}^{(0)},s_{2}^{(0)}\right\} 
$ is computed for $x_{sic}=A_{110}+jA_{110}$ and $r_{2,sic}=%
\alpha_{2}x_{sic}+w_{2}$, 
\begin{multline}
P_{b_{21}}|_{\mathbb{A}_{222}}\mathrm{P\!}\left( \hat{b}%
_{11}^{(0)}|_{b_{11}^{(0)}\text{,}s_{2}^{(0)}}\right) =\mathrm{P\!}\left( 
\mathfrak{n}_{2}+\alpha_{2}A_{110}\geq0\right)  \label{P-Pb-21-01} \\
\times\mathrm{P\!}\left( \mathfrak{n}_{2}+\alpha_{2}A_{010}\leq0\left\vert 
\mathfrak{n}_{2}+\alpha_{2}A_{110}\geq0\right. \right)
\end{multline}
Since the RHS of (\ref{P-Pb-21-01}) can be simplified to $\mathrm{P\!}\left( 
\mathfrak{n}_{2}\leq\alpha_{2}A_{0\acute{1}0},\mathfrak{n}_{2}\geq\alpha
_{2}A_{\acute{1}\acute{1}0}\right) $, then 
\begin{equation}
P_{b_{21}}|_{\mathbb{A}_{222}}\mathrm{P\!}\left( \hat{b}%
_{11}^{(0)}|_{b_{11}^{(0)}\text{,}s_{2}^{(0)}}\right) =Q\left( \sqrt{%
\gamma_{2,1}}\right) -Q\left( \sqrt{\gamma_{2,3}}\right)
\end{equation}
where $\gamma_{2,3}=\alpha_{2}^{2}A_{110}^{2}/\sigma_{\mathfrak{n}_{2}}^{2}$.

For $b_{22}$, the error probability is given by%
\begin{multline}
P_{b_{22}}|_{\mathbb{A}_{222}}\mathrm{P\!}\left( \hat{b}%
_{11}^{(0)}|_{b_{11}^{(0)}\text{,}s_{2}^{(0)}}\right) =\mathrm{P\!}\left( 
\mathfrak{n}_{2}+\alpha_{2}A_{110}\geq0\right)  \label{E-Pb22-02} \\
\times\mathrm{P\!}\left( \mathfrak{q}_{2}+\alpha_{2}A_{010}\leq0\left\vert 
\mathfrak{n}_{2}+\alpha_{2}A_{110}\geq0\right. \right) .
\end{multline}
The RHS of (\ref{E-Pb22-02}) can be simplified to $\mathrm{P\!}\left( 
\mathfrak{q}_{2}\leq\alpha_{2}A_{0\acute{1}0}\right) \mathrm{P\!}\left( 
\mathfrak{n}_{2}\geq\alpha_{2}A_{\acute{1}\acute{1}0}\right) $, and thus%
\begin{equation}
P_{b_{22}}|_{\mathbb{A}_{222}}\mathrm{P\!}\left( \hat{b}%
_{11}^{(0)}|_{b_{11}^{(0)},s_{2}^{(0)}}\right) =Q\left( \sqrt{\gamma_{2,1}}%
\right) \left( 1-Q\left( \sqrt{\gamma_{2,3}}\right) \right) .
\end{equation}

The remaining cases, $\boldsymbol{Case}$ $\boldsymbol{3}$ to $\boldsymbol{%
Case}$ $\boldsymbol{9}$ where the constellation points are $0110$, $0011$, $%
0111$, $1000$, $1001$, $1100$, $1101$ are similar to $\boldsymbol{Case}$ $%
\boldsymbol{1,}$ and $\boldsymbol{Case}$ $\boldsymbol{10}$ to $\boldsymbol{%
Case}$ $\boldsymbol{16}$ where the constellation points are $0001$, $0100$, $%
0101$, $1010$, $1011$, $1110$, $1111$ are similar to $\boldsymbol{Case}$ $%
\boldsymbol{2}$. By using (\ref{Eb21}), the BER of the second user given
that $\hat{b}_{11}=b_{11}$ can be expressed as 
\begin{multline}
P_{U_{2}}^{\left( 1\right) }=\frac{1}{2}Q\left( \sqrt{\gamma_{2,1}}\right) %
\left[ 2-Q\left( \sqrt{\gamma_{2,2}}\right) -Q\left( \sqrt{\gamma_{2,3}}%
\right) \right]  \label{Eb55} \\
-\frac{1}{2}Q\left( \sqrt{\gamma_{2,3}}\right) .
\end{multline}
The constellation diagram of $x_{sic}$, after subtracting $\hat{s}_{1}|_{%
\hat{b}_{11}\neq b_{11}}$ is shown in Fig. \ref{2U_cases_U2} (solid
diamonds). The total error probability of the second user when $\hat{b}%
_{11}\neq b_{11}$ can be derived by considering all the cases in (\ref{Eb21}%
).

$\boldsymbol{Case}$ $\boldsymbol{1}$: $s_{1}^{(0)}$ $\left(
b_{11}^{(0)}\right) $, $\hat{s}_{1}^{(2)}$ $\left( \hat{b}_{11}^{(1)}\right) 
$, $s_{2}^{(2)}$:

The transmitted signal amplitude of this point is $x=A_{1\acute{1}%
0}+jA_{110} $, thus, $x_{I,sic}=A_{2\acute{1}0}$. The error probability for
this case is 
\begin{multline}
P_{b_{21}}|_{\mathbb{B}_{221}}\mathrm{P\!}\left( \hat{b}%
_{11}^{(1)}|_{b_{11}^{(0)}\text{, }s_{2}^{(2)}}\right) =\mathrm{P\!}\left( 
\mathfrak{n}_{2}+\alpha_{2}A_{1\acute{1}0}\leq0\right)  \label{E-Pb-21-03} \\
\times\mathrm{P\!}\left( \mathfrak{n}_{2}+\alpha_{2}A_{2\acute{1}%
0}\geq0\left\vert \mathfrak{n}_{2}+\alpha_{2}A_{1\acute{1}0}\leq0\right.
\right)
\end{multline}
where $\mathbb{B}_{221}\rightarrow\left\{ b_{11}^{(0)},\hat{b}%
_{11}^{(1)},s_{2}^{(2)}\right\} $. The RHS of (\ref{E-Pb-21-03}) can be
simplified to $\mathrm{P\!}\left( \alpha_{2}A_{\acute{2}10}\leq\mathfrak{n}%
_{2}\leq\alpha_{2}A_{\acute{1}10}\right) $. Thus, 
\begin{equation}
P_{b_{21}}|_{\mathbb{B}_{221}}\mathrm{P\!}\left( \hat{b}%
_{11}^{(1)}|_{b_{11}^{(0)}\text{, }s_{2}^{(2)}}\right) =Q\left( \sqrt{%
\gamma_{2,2}}\right) -Q\left( \sqrt{\gamma_{2,4}}\right)
\end{equation}
where $\gamma_{2,4}=\alpha_{2}^{2}A_{2\acute{1}0}^{2}/\sigma_{\mathfrak{n}%
_{2}}^{2}$.

The error probability of $b_{22}$ is obtained as 
\begin{multline}
P_{b_{22}}|_{\mathbb{B}_{221}}\mathrm{P\!}\left( \hat{b}%
_{11}^{(1)}|_{b_{11}^{(0)}\text{, }s_{2}^{(2)}}\right) =\mathrm{P\!}\left( 
\mathfrak{n}_{2}+\alpha_{2}A_{1\acute{1}0}\leq0\right)  \label{E-Pb22-03} \\
\times\mathrm{P\!}\left( \mathfrak{q}_{2}+\alpha_{2}A_{010}\leq0\left\vert 
\mathfrak{n}_{2}+\alpha_{2}A_{1\acute{1}0}\leq0\right. \right)
\end{multline}
By noting that the RHS of (\ref{E-Pb22-03}) can be simplified to $\mathrm{P\!%
}\left( \mathfrak{q}_{2}\leq\alpha_{2}A_{0\acute{1}0},\mathfrak{n}%
_{2}\leq\alpha_{2}A_{\acute{1}10}\right) $, then%
\begin{equation}
P_{b_{22}}|_{\mathbb{B}_{221}}\mathrm{P\!}\left( \hat{b}%
_{11}^{(1)}|_{b_{11}^{(0)}\text{, }s_{2}^{(2)}}\right) =Q\left( \sqrt{%
\gamma_{2,1}}\right) Q\left( \sqrt{\gamma_{2,2}}\right) .
\end{equation}

$\boldsymbol{Case}$ $\boldsymbol{2}$: $s_{1}^{(0)}$ $\left(
b_{11}^{(0)}\right) $, $\hat{s}_{1}^{(2)}$ $\left( \hat{b}_{11}^{(1)}\right) 
$, $s_{2}^{(0)}$:

The error probability of this case can be computed as%
\begin{multline}
P_{b_{21}}|_{\mathbb{B}_{222}}\mathrm{P\!}\left( \hat{b}%
_{11}^{(1)}|_{b_{11}^{(0)}\text{,}s_{2}^{(0)}}\right) =\mathrm{P\!}\left( 
\mathfrak{n}_{2}+\alpha_{2}A_{110}\leq0\right)  \label{E-Pb21-04} \\
\times\mathrm{P\!}\left( \mathfrak{n}_{2}+\alpha_{2}A_{210}\leq0\left\vert 
\mathfrak{n}_{2}+\alpha_{2}A_{110}\leq0\right. \right)
\end{multline}
The RHS of (\ref{E-Pb21-04}) can be expressed as $\mathrm{P\!}\left( 
\mathfrak{n}_{2}\leq\alpha_{2}A_{\acute{2}\acute{1}0},\mathfrak{n}%
_{2}\leq\alpha_{2}A_{\acute{1}\acute{1}0}\right) $, and hence,%
\begin{equation}
P_{b_{21}}|_{\mathbb{B}_{222}}\mathrm{P\!}\left( \hat{b}%
_{11}^{(1)}|_{b_{11}^{(0)}\text{,}s_{2}^{(0)}}\right) =Q\left( \sqrt{%
\gamma_{2,5}}\right)
\end{equation}
\bigskip where $\gamma_{2,5}=\alpha_{2}^{2}A_{210}^{2}/\sigma_{\mathfrak{n}%
_{2}}^{2}$, and $\mathbb{B}_{222}\rightarrow\left\{ b_{11}^{(0)},\hat{b}%
_{11}^{(1)},s_{2}^{(0)}\right\} .$

For $b_{22}$, the error probability is evaluated as 
\begin{multline}
P_{b_{22}}|_{\mathbb{B}_{222}}\mathrm{P\!}\left( \hat{b}%
_{11}^{(1)}|_{b_{11}^{(0)}\text{,}s_{2}^{(0)}}\right) =\mathrm{P\!}\left( 
\mathfrak{n}_{2}+\alpha_{2}A_{110}\leq0\right)  \label{E-Pb22-05} \\
\times\mathrm{P\!}\left( \mathfrak{q}_{2}+\alpha_{2}A_{010}\leq0\left\vert 
\mathfrak{n}_{2}+\alpha_{2}A_{110}\leq0\right. \right)
\end{multline}
Since the RHS\ of (\ref{E-Pb22-05}) can be simplified to $\mathrm{P\!}\left( 
\mathfrak{q}_{2}\leq\alpha_{2}A_{0\acute{1}0},\mathfrak{n}_{2}\leq\alpha
_{2}A_{\acute{1}\acute{1}0}\right) $, then 
\begin{equation}
P_{b_{22}}|_{\mathbb{B}_{222}}\mathrm{P\!}\left( \hat{b}%
_{11}^{(1)}|_{b_{11}^{(0)}\text{,}s_{2}^{(0)}}\right) =Q\left( \sqrt{%
\gamma_{2,1}}\right) Q\left( \sqrt{\gamma_{2,3}}\right) .
\end{equation}

For the remaining cases, $\boldsymbol{Case}$ $\boldsymbol{3}$ to $%
\boldsymbol{Case}$ $\boldsymbol{9}$ where the associated symbols are $0011$, 
$0110$, $0111$, $1000$, $1001$, $1100$, and $1101$ are identical to $%
\boldsymbol{Case}$ $\boldsymbol{1}$. $\boldsymbol{Case}$ $\boldsymbol{%
\boldsymbol{1}0}$ to $\boldsymbol{Case}$ $\boldsymbol{\boldsymbol{1}6}$
where the constellation points are $0001$, $0100$, $0101$, $1010$, $1011$, $%
1110$, and $1111$ are similar to $\boldsymbol{Case}$ $\boldsymbol{2}$.
Therefore, the total error probability when $\hat{b}_{11}$ $\neq b_{11}$can
be expressed as%
\begin{multline}
P_{U_{2}}^{\left( 2\right) }=Q\left( \sqrt{\gamma_{2,1}}\right) \left[
Q\left( \sqrt{\gamma_{2,2}}\right) +Q\left( \sqrt{\gamma_{2,3}}\right) %
\right] +Q\left( \sqrt{\gamma_{2,2}}\right)  \label{Eb66} \\
-Q\left( \sqrt{\gamma_{2,4}}\right) +Q\left( \sqrt{\gamma_{2,5}}\right) .
\end{multline}
Finally, the exact total BER for the second user is given as the sum of the
results of the two scenarios where $\hat{b}_{11}=b_{11}$ and $\hat{b}%
_{11}\neq b_{11}$,%
\begin{equation}
P_{U_{2}}=\frac{1}{2}\sum_{i=1}^{5}v_{i}Q\left( \sqrt{\gamma_{2,i}}\right) 
\text{, }\mathbf{v}=\left[ 2,1,-1,-1,1\right] .  \label{coc22}
\end{equation}

\subsection{Average BER, $N=2$\label{SubSec-averageBER U2}}

The average BER can be evaluated by averaging over the PDFs of all $%
\gamma_{n,c}$ values, which are given in Appendix I. Therefore, by
substituting $N=2$ and $n=\left[ 1,2\right] $ in the ordered PDF in \ (\ref%
{E-Ordered-PDF}), the exact average BER of the first and second users can be
simplified to%
\begin{multline}
\overline{P}_{U_{1}}=\frac{1}{\pi\Gamma(m)}\sum_{c=1}^{2}\sum_{k=0}^{2}%
\sum_{i=0}^{\infty}\left( -1\right) ^{k}S_{i}\left( \frac{m}{\overline {%
\gamma}_{1,c}}\right) ^{m(1+k)}  \label{Naka_2U_1} \\
\times\int_{0}^{\frac{\pi}{2}}\frac{\left( i+mk\right) !}{\left( \frac {1}{%
2\sin^{2}\left( \psi_{1,c}\right) }+\frac{m\text{ }\left( 1+k\right) }{%
\overline{\gamma}_{1,c}}\right) ^{i+mk+1}}d\psi_{n,c}
\end{multline}
and%
\begin{multline}
\bar{P}_{U_{2}}=\frac{1}{\pi\Gamma(m)}\sum\limits_{c=1}^{5}\sum_{i=0}^{%
\infty }v_{c}S_{i}\left( \frac{m}{\overline{\gamma}_{2,c}}\right) ^{2m}
\label{Naka_2U_2} \\
\times\int_{0}^{\frac{\pi}{2}}\frac{\left( i+m\right) !}{\left( \frac {1}{%
2\sin^{2}\left( \psi_{2,c}\right) }+\frac{2m\text{ }}{\overline{\gamma }%
_{2,c}}\right) ^{i+m+1}}d\psi_{2,c}
\end{multline}
where $\mathbf{v}=\left[ 2,1,-1,-1,1\right] $. It is interesting to note
that for the special case of Rayleigh fading channel where $m=1$, the BER
for both users can be expressed in closed-form as, 
\begin{equation}
\bar{P}_{U_{1}}=\frac{1}{4}\sum_{c=1}^{2}\left( 1-\frac{1}{\sqrt{\frac {1}{%
\overline{\gamma}_{1,c}}+1}}\right)  \label{Pe11}
\end{equation}
and%
\begin{equation}
\bar{P}_{U_{2}}=\frac{1}{2}\sum_{c=1}^{5}v_{c}\left( \sqrt{\frac {\overline{%
\gamma}_{2,c}}{\overline{\gamma}_{2,c}+1}}-\sqrt{\frac {8\overline{\gamma}%
_{2,c}}{2\overline{\gamma}_{2,c}+1}}+1\right)  \label{Pe22}
\end{equation}
where $\mathbf{v}=\left[ 2,1,-1,-1,1\right] $.

\section{\label{SubSec-BER U1-3U}NOMA Bit Error Rate (BER) Analysis: Three
Users ($N=3)$}

\begin{figure*}[tb]
    \centering
    \includegraphics{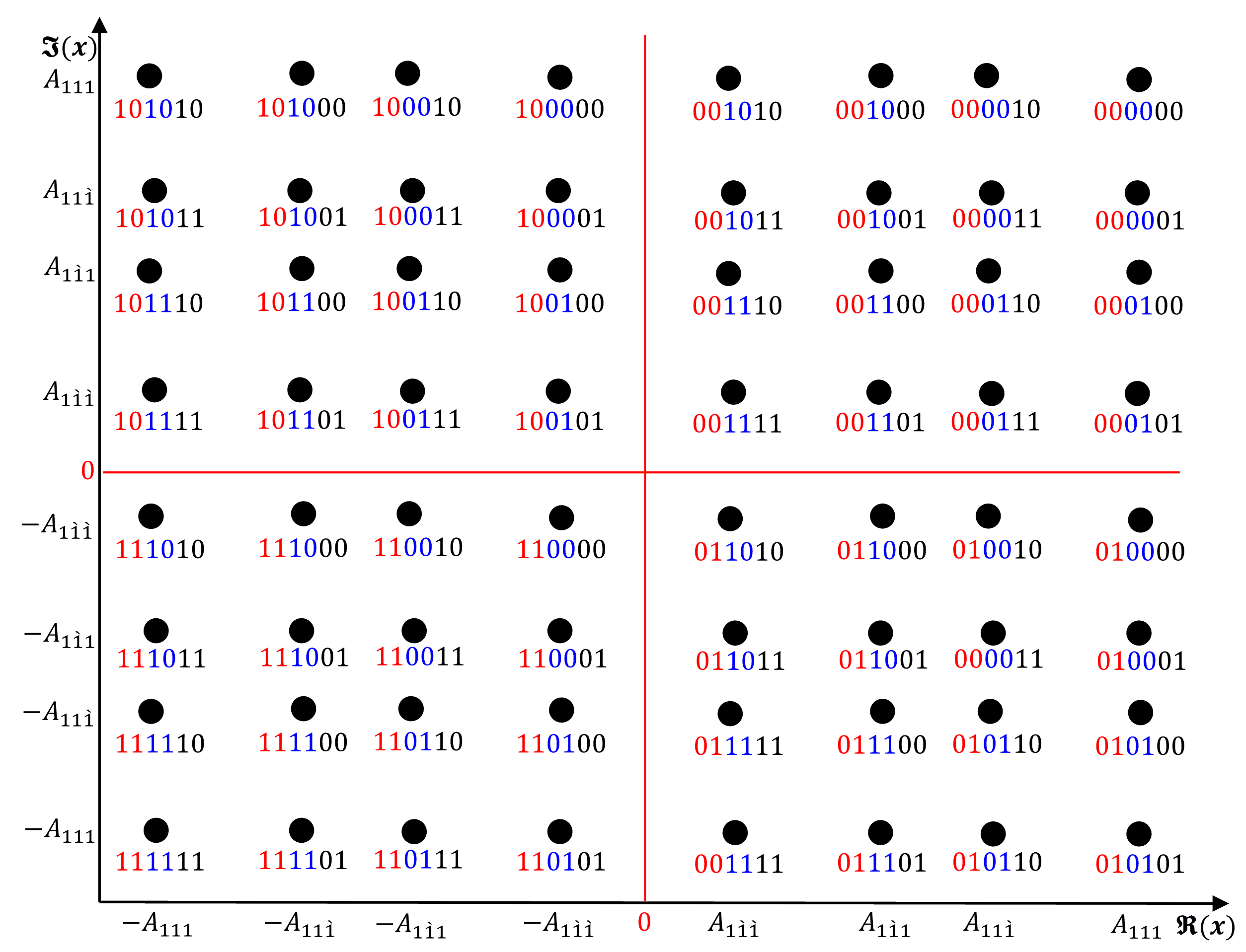}
    \caption{The transmitted superimposed signal constellation for $N=3$.}
    \label{3u_fig}
\end{figure*}

This section presents the derivation of the BER for a three-users NOMA
system, $N=3$, which generally follows the derived $N=2$ case. The
transmitted signal constellation is given in Fig. \ref{3u_fig}. The first,
second, and third users' signals are shown in the form of $\left[
s_{1},s_{2},s_{3}\right] $.\ The binary bit representation for the three
users are represented as $%
\begin{bmatrix}
b_{11} & b_{12} & b_{21} & b_{22} & b_{31} & b_{32}%
\end{bmatrix}
$, for each bit $b_{ni}$, $n=\left[ 1\text{, }2,3\right] $, and $i=\left[ 1,2%
\right] $ which denotes bits' indices. 


\subsection{BER of First User $\left( U_{1}|_{N=3}\right) $}

The error probability of the first user can be obtained directly from Fig. %
\ref{3u_fig}. The probability of error for each bit depends on the values of 
$s_{1}$, $s_{2}$ and $s_{3}$. For example, given that $s_{1}^{\left(
2\right) }$, $s_{3}^{(3)}$, the first bit $b_{11}$ might be detected
incorrectly if $\hat{s}_{1}=a_{0}$ $(00)$ or $a_{1}$ $(01)$, as shown in
Fig. \ref{3u_fig}. Therefore, the average BER should consider all possible
combinations of $s_{1}$, $s_{2}$ and $s_{3}$,%
\begin{equation}
P_{b_{1i}}=\sum_{c,k,v}P_{b_{1i}}|_{s_{1}^{(c)},s_{2}^{(k)},s_{3}^{(v)}}%
\mathrm{P\!}\left( s_{1}^{(c)},s_{2}^{(k)},s_{3}^{(v)}\right) .
\label{Eb_321}
\end{equation}
Because $s_{1}$, $s_{2}$ and $s_{3}$ are mutually independent, (\ref{Eb_321}%
) can be written as 
\begin{equation}
P_{b_{1i}}=\frac{1}{64}\sum_{\left\{ c,k,v\right\}
=0}^{3}P_{b_{1i}}|_{s_{1}^{(c)},s_{2}^{(k)},s_{3}^{(v)}}.
\end{equation}
The error probability of the first user is derived by considering all
possible combinations, which can be derived as follows:

$\boldsymbol{Case}$ $\boldsymbol{1}$: $s_{1}^{(0)}$, $s_{2}^{(2)}$, $%
s_{3}^{(2)}$:

For this case, $x=A_{1\acute{1}\acute{1}}+jA_{111}$, consequently, the error
probability of $b_{11}$ is given by,%
\begin{equation}
P_{b_{1i}}|_{s_{1}^{(0)},s_{2}^{(2)},s_{3}^{(2)}}=\mathrm{P\!}\left( \hat {s}%
_{1}=a_{2};a_{3}\right)
\end{equation}
As can be noted from Fig. \ref{3u_fig}, $\mathrm{P\!}\left( \hat{s}%
_{1}=a_{2};a_{3}\right) $ depends only on the inphase component of $\check{r}%
_{1}$, i.e., $\mathfrak{r}_{1}=\alpha_{1}A_{1\acute{1}\acute{1}}+\mathfrak{n}%
_{1}$, and the specific value of $x$. Thus,%
\begin{align}
P_{b_{1i}}|_{s_{1}^{(0)},s_{2}^{(2)},s_{3}^{(2)}} & =\mathrm{P\!}\left( 
\mathfrak{r}_{1}\leq0\right)  \notag \\
& =\mathrm{P\!}\left( \alpha_{1}A_{1\acute{1}\acute{1}}+\mathfrak{n}%
_{1}\leq0\right)  \notag \\
& =Q\left( \sqrt{\gamma_{3,1}}\right) .
\end{align}
Following the same approach, Table \ref{Table-cases-U1-N3} shows the summary
of $\boldsymbol{Case}$ $\mathbf{2}$ to $\boldsymbol{Case}$ $\mathbf{4}$%
\textbf{. }The remaining $60$ cases, $\boldsymbol{Case}$ $\boldsymbol{5}$ to 
$\boldsymbol{64}$ in Fig. \ref{3u_fig}, can be obtained following the same
approach, and hence, the total BER of the first user can be expressed as%
\begin{equation}
P_{U_{1}}=\frac{1}{4}\sum\limits_{v=1}^{4}Q\left( \sqrt{\gamma_{3,v}}\right)
.
\end{equation}

\begin{table}[tbp] \centering%
\caption{Summary of cases 1 to 4 for $U_{1}$, $N=3.$}%
\begin{tabular}{|c|c|c|c|}
\hline
$Case$ $(v)$ & $x$ & $s_{1},s_{2},s_{3}$ & $\gamma_{3,v}$ \\ \hline\hline
$1$ & $A_{1\acute{1}\acute{1}}+jA_{111}$ & \multicolumn{1}{|l|}{$%
a_{0},a_{2},a_{2}$} & \multicolumn{1}{|l|}{$\frac{\alpha_{1}^{2}A_{1\acute{1}%
\acute{1}}^{2}}{\sigma_{\mathfrak{n}_{1}}^{2}}$} \\ \hline
$2$ & $A_{1\acute{1}1}+jA_{111}$ & \multicolumn{1}{|l|}{$a_{0},a_{2},a_{0}$}
& \multicolumn{1}{|l|}{$\frac{\alpha_{1}^{2}A_{1\acute{1}1}^{2}}{\sigma _{%
\mathfrak{n}_{1}}^{2}}$} \\ \hline
$3$ & $A_{11\acute{1}}+jA_{111}$ & \multicolumn{1}{|l|}{$a_{0},a_{0},a_{2}$}
& \multicolumn{1}{|l|}{$\frac{\alpha_{1}^{2}A_{11\acute{1}}^{2}}{\sigma _{%
\mathfrak{n}_{1}}^{2}}$} \\ \hline
$4$ & $A_{111}+jA_{111}$ & $a_{0},a_{0},a_{0}$ & $\frac{%
\alpha_{1}^{2}A_{111}^{2}}{\sigma_{\mathfrak{n}_{1}}^{2}}$ \\ \hline
\end{tabular}
\label{Table-cases-U1-N3}%
\end{table}%


\begin{figure*}[tb]
    \centering
    \includegraphics[width=6.5in]{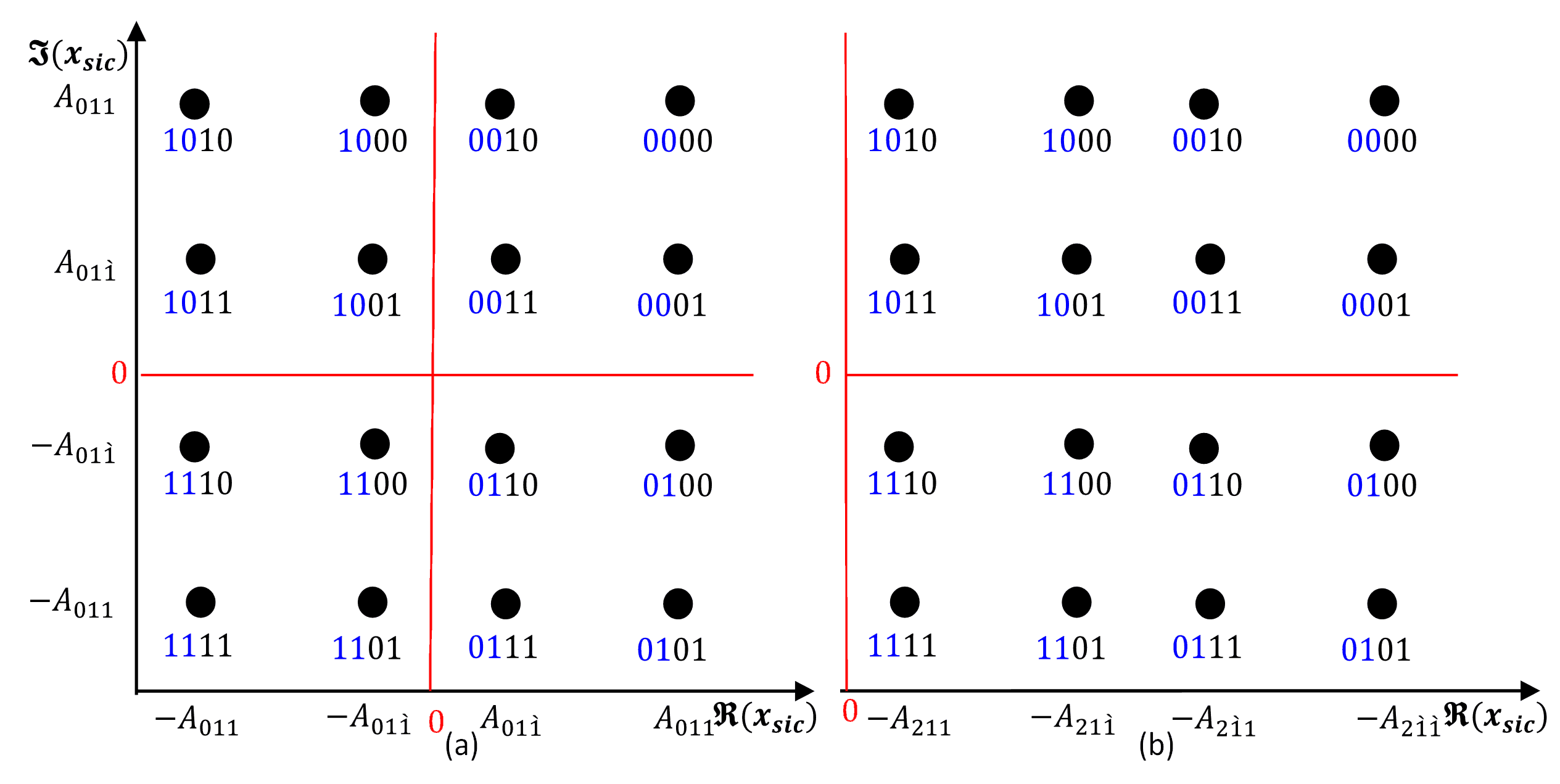}
    \caption{Equivelant constellation of (a) $x_{sic}|_{\hat{b}_{11}=b_{11}}$ and (b) $x_{sic}|_{\hat{b}_{11}\neq b_{11}}$ of the  second user, $N=3$.}
    \label{user2_3u}
\end{figure*}

\subsection{BER of Second User $\left( U_{2}|_{N=3}\right) $}

The BER of the second user depends on the detection result of the first user
and the SIC process as illustrated in Fig. \ref{user2_3u}. The first case is
when $\hat{s}_{1}=s_{1}$, which is represented in Fig. \ref{user2_3u}.a
while the other case is when $\hat{s}_{1}\neq s_{1}$, which is depicted in
Fig. \ref{user2_3u}.b. To detect its own symbol $s_{2}$, the second user
should follow the SIC process described in (\ref{E-13}). It should be noted
that the error probability of the second user is not affected by the
detection result of the second bit of the first user $b_{12}$.

The BER of $U_{2}$ depends on $s_{1}$, $\hat{s}_{1}$, $s_{2}$ and 
$s_{3}$. Therefore, the average BER for $b_{2i}$ is the average of all
possible combinations,%
\begin{align}
P_{b_{2i}} & =\sum_{g,l,k,v}P_{b_{2i}}|_{s_{1}^{(g)},s_{2}^{(k)},s_{3}^{(v)},%
\hat{s}_{1}^{(l)}}\mathrm{P\!}\left( s_{1}^{(g)},s_{2}^{(k)},s_{3}^{(v)},%
\hat{s}_{1}^{(l)}\right)  \notag \\
& =\sum_{g,l,k,v}P_{b_{2i}}|_{s_{1}^{(g)},s_{2}^{(k)},s_{3}^{(v)},\hat{s}%
_{1}^{(l)}}\mathrm{P\!}\left( \hat{s}%
_{1}^{(l)}|_{s_{1}^{(g)},s_{2}^{(k)},s_{3}^{(v)}}\right)  \notag \\
& \text{ \ \ \ \ \ \ \ \ \ \ \ \ \ \ \ \ \ \ \ \ \ \ \ \ \ \ \ \ \ \ \ \ \ }%
\times\mathrm{P\!}\left( s_{1}^{(g)},s_{2}^{(k)},s_{3}^{(v)}\right)  \notag
\\
& =\frac{1}{64}\sum_{\left\{ g,l,k,v\right\}
=0}^{3}P_{b_{2i}}|_{s_{1}^{(g)},s_{2}^{(k)},s_{3}^{(v)},\hat{s}_{1}^{(l)}} 
\notag \\
& \text{ \ \ \ \ \ \ \ \ \ \ \ \ \ \ \ \ \ \ \ \ \ \ \ \ \ \ \ \ \ \ \ \ }%
\times\mathrm{P\!}\left( \hat{s}%
_{1}^{(l)}|_{s_{1}^{(g)},s_{2}^{(k)},s_{3}^{(v)}}\right)
\end{align}
With the aid of Fig. \ref{user2_3u}.a, the total BER for the scenario where $%
\hat{s}_{1}=s_{1}$, or more specifically, where $\hat{b}_{11}=b_{11}$ is
obtained as follows:

$\boldsymbol{Case}$ $\boldsymbol{1}$: $s_{1}^{(0)}$ $\left(
b_{11}^{(0)}\right) $, $\hat{s}_{1}^{(0)}$ $\left( \hat{b}_{11}^{(0)}\right) 
$, $s_{2}^{(2)}$, $s_{3}^{(2)}$:

The probability $P_{b_{21}}|_{\mathbb{A}_{321}}$, $\mathbb{A}%
_{321}\rightarrow\left\{ b_{11}^{(0)},\hat{b}%
_{11}^{(0)},s_{2}^{(2)},s_{3}^{(2)}\right\} $ can be evaluated by
considering the error probability of $U_{1}$ given $\mathbb{A}_{321}$. In
this case $P_{b_{21}}$ can be computed by noting that $P_{b_{21}}|_{\mathbb{A%
}_{321}}=\mathrm{P\!}\left( \hat{s}_{2}=a_{0};a_{1}|_{b_{11}^{(0)}\text{, }%
\hat{b}_{11}^{(0)}\text{, }s_{2}^{(2)}\text{, }s_{3}^{(2)}}\right) $. The
transmitted signal amplitude of this case is $x=A_{1\acute{1}\acute{1}%
}+jA_{111}$, thus, $x_{sic}=A_{0\acute{1}\acute{1}}+jA_{011}$ and $%
r_{2,sic}=\alpha_{2}x_{sic}+w_{2}$. Therefore, 
\begin{align}
P_{b_{21}}|_{\mathbb{A}_{321}} & =\mathrm{P\!}\left( \mathfrak{n}%
_{2}+\alpha_{2}A_{0\acute{1}\acute{1}}\geq0\left\vert \mathfrak{n}_{2}+A_{1%
\acute{1}\acute{1}}\geq0\right. \right)  \notag \\
& =\mathrm{P\!}\left( \mathfrak{n}_{2}\geq\alpha_{2}A_{011}\left\vert 
\mathfrak{n}_{2}\geq\alpha_{2}A_{\acute{1}11}\right. \right)  \label{Eb111}
\end{align}
where $\mathfrak{r}_{2,sic}=\alpha_{2}A_{0\acute{1}\acute{1}}+\mathfrak{n}%
_{2}$. Therefore, 
\begin{align}
P_{b_{21}}|_{\mathbb{A}_{321}}\mathrm{P\!}\left( \hat{b}%
_{11}^{(0)}|_{b_{11}^{(0)}\text{,}s_{2}^{(2)},s_{3}^{(2)}}\right) & =\mathrm{%
P\!}\left( \mathfrak{n}_{2}\geq\alpha_{2}A_{011}\right)  \notag \\
& =Q\left( \sqrt{\gamma_{3,5}}\right)
\end{align}
where $\gamma_{3,5}=\alpha_{2}^{2}A_{011}^{2}/\sigma_{\mathfrak{n}_{2}}^{2}$.

For $b_{22\text{ }}$bit, the error probability is obtained in a similar
approach as $b_{21}$ 
\begin{align}
P_{b_{22}}|_{\mathbb{A}_{321}}\mathrm{P\!}\left( \hat{b}%
_{11}^{(0)}|_{b_{11}^{(0)}\text{,}s_{2}^{(2)},s_{3}^{(2)}}\right) & =\mathrm{%
P\!}\left( \mathfrak{q}_{2}\leq\alpha_{2}A_{0\acute{1}\acute{1}}\right) 
\mathrm{P\!}\left( \mathfrak{n}_{2}\geq\alpha_{2}A_{\acute{1}11}\right) 
\notag \\
& =Q\left( \sqrt{\gamma_{3,5}}\right) \left( 1-Q\left( \sqrt{\gamma _{3,1}}%
\right) \right) .
\end{align}

$\boldsymbol{Case}$ $\boldsymbol{2}$: $s_{1}^{(0)}$ $\left(
b_{11}^{(0)}\right) $, $\hat{s}_{1}^{(0)}$ $\left( \hat{b}_{11}^{(0)}\right) 
$, $s_{2}^{(2)}$, $s_{3}^{(0)}$:

Based on Fig. \ref{user2_3u}.a, the error probability of $b_{21}$ is given by%
\begin{multline}
P_{b_{21}}|_{\mathbb{A}_{321}}\mathrm{P\!}\left( \hat{b}%
_{11}^{(0)}|_{b_{11}^{(0)}\text{,}s_{2}^{(2)},s_{3}^{(0)}}\right) =\mathrm{%
P\!}\left( \mathfrak{n}_{2}\geq\alpha_{2}A_{\acute{1}1\acute{1}}\right)
\label{E-Pb22-06} \\
\times\mathrm{P\!}\left( \mathfrak{n}_{2}\geq\alpha_{2}A_{01\acute{1}%
}\left\vert \mathfrak{n}_{2}\geq\alpha_{2}A_{\acute{1}1\acute{1}}\right.
\right)
\end{multline}
By noting that the RHS of (\ref{E-Pb22-06}) can be simplified to $\mathrm{P\!%
}\left( \mathfrak{n}_{2}\geq\alpha_{2}A_{01\acute{1}}\right) $, and
consequently it is straightforward to show that 
\begin{equation}
P_{b_{21}}|_{\mathbb{A}_{322}}\mathrm{P\!}\left( \hat{b}%
_{11}^{(0)}|_{b_{11}^{(0)}\text{,}s_{2}^{(2)},s_{3}^{(0)}}\right) =Q\left( 
\sqrt {\gamma_{3,6}}\right)
\end{equation}
where $\gamma_{3,6}=\alpha_{2}^{2}A_{01\acute{1}}^{2}/\sigma_{\mathfrak{n}%
_{2}}^{2}$, $\mathbb{A}_{322}\rightarrow\left\{ b_{11}^{\left( 0\right) },%
\hat{b}_{11}^{(0)},s_{2}^{(2)},s_{3}^{(0)}\right\} $.

For $b_{22\text{ }}$bit, the error probability is obtained as 
\begin{align}
P_{b_{22}}|_{\mathbb{A}_{322}}\mathrm{P\!}\left( \hat{b}%
_{11}^{(0)}|_{b_{11}^{(0)}\text{,}s_{2}^{(2)},s_{3}^{(0)}}\right) & =\mathrm{%
P\!}\left( \mathfrak{q}_{2}\leq\alpha_{2}A_{0\acute{1}\acute{1}}\right) 
\mathrm{P\!}\left( \mathfrak{n}_{2}\geq\alpha_{2}A_{\acute{1}1\acute{1}%
}\right)  \notag \\
& =Q\left( \sqrt{\gamma_{3,5}}\right) \left( 1-Q\left( \sqrt{\gamma _{3,2}}%
\right) \right) .
\end{align}

$\boldsymbol{Case}$ $\boldsymbol{3}$: $s_{1}^{(0)}$ $\left(
b_{11}^{(0)}\right) $, $\hat{s}_{1}^{(0)}$ $\left( \hat{b}_{11}^{(0)}\right) 
$, $s_{2}^{(0)}$, $s_{3}^{(2)}$:

The error probability of this case can be evaluated as%
\begin{align}
P_{b_{21}}|_{\mathbb{A}_{323}}\mathrm{P\!}\left( \hat{b}%
_{11}^{(0)}|_{b_{11}^{(0)},s_{2}^{(0)},s_{3}^{(2)}}\right) & =\mathrm{P\!}%
\left( \alpha_{2}A_{\acute{1}\acute{1}1}\leq\mathfrak{n}_{2}\leq%
\alpha_{2}A_{0\acute{1}1}\right)  \notag \\
& =Q\left( \sqrt{\gamma_{3,6}}\right) -Q\left( \sqrt{\gamma_{3,3}}\right) .
\end{align}
where $\mathbb{A}_{323}\rightarrow\left\{ b_{11}^{(0)},\hat{b}%
_{11}^{(0)},s_{2}^{(0)},s_{3}^{(2)}\right\} .$

For $b_{22\text{ }}$bit, the error probability is obtained as%
\begin{align}
P_{b_{22}}|_{\mathbb{A}_{323}}\mathrm{P\!}\left( \hat{b}%
_{11}^{(0)}|_{b_{11}^{(0)},s_{2}^{(0)},s_{3}^{(2)}}\right) & =\mathrm{P\!}%
\left( \mathfrak{q}_{2}\leq\alpha_{2}A_{0\acute{1}\acute{1}}\right)  \notag
\\
& \text{ \ \ \ \ \ \ \ \ \ \ \ \ }\times\mathrm{P\!}\left( \mathfrak{n}%
_{2}\geq\alpha_{2}A_{\acute{1}\acute{1}1}\right)  \notag \\
& =Q\left( \sqrt{\gamma_{3,5}}\right) \left( 1-Q\left( \sqrt{\gamma _{3,3}}%
\right) \right) .
\end{align}

$\boldsymbol{Case}$ $\boldsymbol{4}$: $s_{1}^{(0)}$ $\left(
b_{11}^{(0)}\right) $, $\hat{s}_{1}^{(0)}$ $\left( \hat{b}_{11}^{(0)}\right) 
$, $s_{2}^{(0)}$, $s_{3}^{(0)}$:

The error probability of $b_{21}$ can be evaluated as%
\begin{align}
P_{b_{21}}|_{\mathbb{A}_{324}}\mathrm{P\!}\left( \hat{b}%
_{11}^{(0)}|_{b_{11}^{(0)}\text{,}s_{2}^{(0)},s_{3}^{(0)}}\right) & =\mathrm{%
P\!}\left( \alpha_{2}A_{\acute{1}\acute{1}\acute{1}}\leq\mathfrak{n}%
_{2}\leq\alpha_{2}A_{0\acute{1}\acute{1}}\right)  \notag \\
& =Q\left( \sqrt{\gamma_{3,5}}\right) -Q\left( \sqrt{\gamma_{3,4}}\right) .
\end{align}
where $\mathbb{A}_{324}\rightarrow\left\{ b_{11}^{(0)},\hat{b}%
_{11}^{(0)},s_{2}^{(0)},s_{3}^{(0)}\right\} $.

For $b_{22\text{ }}$, the error probability can be expressed as 
\begin{align}
P_{b_{22}}|_{\mathbb{A}_{324}}\mathrm{P\!}\left( \hat{b}%
_{11}^{(0)}|_{b_{11}^{(0)}\text{,}s_{2}^{(0)},s_{3}^{(0)}}\right) & =\mathrm{%
P\!}\left( \mathfrak{q}_{2}\leq\alpha_{2}A_{0\acute{1}\acute{1}}\right) 
\mathrm{P\!}\left( \mathfrak{n}_{2}\geq\alpha_{2}A_{\acute{1}\acute{1}\acute{%
1}}\right)  \notag \\
& =Q\left( \sqrt{\gamma_{3,5}}\right) \left( 1-Q\left( \sqrt{\gamma _{3,4}}%
\right) \right) .
\end{align}
By taking into account the remaining $60$ cases from $\boldsymbol{Case}$ $%
\boldsymbol{5}$ to $\boldsymbol{64}$, the total BER of the second user when $%
\hat{b}_{11}=$ $b_{11}$\ can be expressed as%
\begin{multline}
P_{U_{2}}^{(1)}=\frac{1}{4}Q\left( \sqrt{\gamma_{3,5}}\right) \left(
6-\sum\limits_{v=1}^{4}Q\left( \sqrt{\gamma_{3,v}}\right) \right) \\
+\frac{1}{4}\sum\limits_{i}d_{i}Q\left( \sqrt{\gamma_{3,i}}\right) \text{, }%
i\in\left\{ 3,4,6\right\} \text{, }\mathbf{d}=\left[ -1,-1,2\right] .
\label{EB2_111}
\end{multline}
The same approach is adopted for the scenario where $\hat{b}_{11}\neq b_{11}$%
. The transmitted signal constellation after subtracting $\hat{s}_{1}$ when $%
\hat{b}_{11}\neq b_{11}$ is shown in Fig. \ref{user2_3u} (b).

$\boldsymbol{Case}$ $\boldsymbol{1}$: $s_{1}^{(0)}$ $\left(
b_{11}^{(0)}\right) $, $\hat{s}_{1}^{(2)}$ $\left( \hat{b}_{11}^{(1)}\right) 
$, $s_{2}^{(2)}$, $s_{3}^{(2)}$:

The transmitted signal in this case is $x=A_{1\acute{1}\acute{1}}+jA_{111}$,
which after subtracting $\hat{s}_{1}=-\sqrt{\beta_{1}}+j\sqrt{\beta_{1}}$
becomes $x_{sic}=A_{2\acute{1}\acute{1}}+jA_{011}$. The error probability
can be computed as 
\begin{align}
P_{b_{21}}|_{\mathbb{B}_{321}}\mathrm{P\!}\left( \hat{b}%
_{11}^{(1)}|_{b_{11}^{(0)}\text{,}s_{2}^{(2)},s_{3}^{(2)}}\right) & =\mathrm{%
P\!}\left( \alpha_{2}A_{\acute{2}11}\leq\mathfrak{n}_{2}\leq\alpha_{2}A_{%
\acute{1}11}\right) \\
& =Q\left( \sqrt{\gamma_{3,1}}\right) -Q\left( \sqrt{\gamma_{3,7}}\right) 
\notag
\end{align}
where $\gamma_{3,7}=\alpha_{2}^{2}A_{2\acute{1}\acute{1}}^{2}/\sigma _{%
\mathfrak{n}_{2}}^{2}$ and $\mathbb{B}_{321}\rightarrow\left\{ b_{11}^{(0)},%
\hat{b}_{11}^{(1)},s_{2}^{(2)},s_{3}^{(2)}\right\} $

For $b_{22}$ bit, 
\begin{align}
P_{b_{22}}|_{\mathbb{B}_{321}}\mathrm{P\!}\left( \hat{b}%
_{11}^{(1)}|_{b_{11}^{(0)}\text{,}s_{2}^{(2)},s_{3}^{(2)}}\right) & =\mathrm{%
P\!}\left( \mathfrak{q}_{2}\leq\alpha_{2}A_{0\acute{1}\acute{1}}\text{, }%
\mathfrak{n}_{2}\leq\alpha_{2}A_{\acute{1}11}\right)  \notag \\
& =Q\left( \sqrt{\gamma_{3,5}}\right) Q\left( \sqrt{\gamma_{3,1}}\right) .
\end{align}

$\boldsymbol{Case}$ $\boldsymbol{2}$: $s_{1}^{(0)}$ $\left(
b_{11}^{(0)}\right) $, $\hat{s}_{1}^{(2)}$ $\left( \hat{b}_{11}^{(1)}\right) 
$, $s_{2}^{(2)}$, $s_{3}^{(0)}$:

The error probability\ of $b_{21}$ for this case can be derived as, 
\begin{align}
P_{b_{21}}|_{\mathbb{B}_{322}}\mathrm{P\!}\left( \hat{b}%
_{11}^{(1)}|_{b_{11}^{(0)}\text{,}s_{2}^{(2)},s_{3}^{(0)}}\right) & =\mathrm{%
P\!}\left( \alpha_{2}A_{\acute{2}1\acute{1}}\leq\mathfrak{n}_{2}\leq\alpha
_{2}A_{\acute{1}1\acute{1}}\right)  \notag \\
& =Q\left( \sqrt{\gamma_{3,2}}\right) -Q\left( \sqrt{\gamma_{3,8}}\right)
\end{align}
where $\gamma_{3,8}=\alpha_{2}^{2}A_{2\acute{1}1}^{2}/\sigma_{\mathfrak{n}%
_{2}}^{2}$ and $\mathbb{B}_{322}\rightarrow\left\{ b_{11}^{(0)}\text{, }\hat{%
b}_{11}^{(1)}\text{, }s_{2}^{(2)}\text{, }s_{3}^{(0)}\right\} $.

The second bit $b_{22}$ error probability can be represented as%
\begin{align}
P_{b_{22}}|_{\mathbb{B}_{322}}\mathrm{P\!}\left( \hat{b}%
_{11}^{(1)}|_{b_{11}^{(0)}\text{,}s_{2}^{(2)},s_{3}^{(0)}}\right) & =\mathrm{%
P\!}\left( \mathfrak{q}_{2}\leq\alpha_{2}A_{0\acute{1}\acute{1}}\right) 
\mathrm{P\!}\left( \mathfrak{n}_{2}\leq\alpha_{2}A_{\acute{1}1\acute{1}%
}\right)  \notag \\
& =Q\left( \sqrt{\gamma_{3,5}}\right) Q\left( \sqrt{\gamma_{3,2}}\right) .
\end{align}
$\boldsymbol{Case}$ $\boldsymbol{3}$: $s_{1}^{(0)}$ $\left(
b_{11}^{(0)}\right) $, $\hat{s}_{1}^{(2)}$ $\left( \hat{b}_{11}^{(1)}\right) 
$, $s_{2}^{(0)}$, $s_{3}^{(2)}$:

Similar to the previous cases, the error probability for this case can be
derived as follows. 
\begin{align}
P_{b_{21}}|_{\mathbb{B}_{323}}\mathrm{P\!}\left( \hat{b}%
_{11}^{(1)}|_{b_{11}^{(0)}\text{,}s_{2}^{(0)},s_{3}^{(2)}}\right) & =\mathrm{%
P\!}\left( \mathfrak{n}_{2}\leq\alpha_{2}A_{\acute{2}\acute{1}1}\right) 
\notag \\
& =Q\left( \sqrt{\gamma_{3,9}}\right)
\end{align}
where $\gamma_{3,9}=\alpha_{2}^{2}A_{21\acute{1}}^{2}/\sigma_{\mathfrak{n}%
_{2}}^{2}$ and $\mathbb{B}_{323}\rightarrow\left\{ b_{11}^{(0)},\hat{b}%
_{11}^{(1)},s_{2}^{(0)},s_{3}^{(2)}\right\} $.

The error probability of $b_{22}$ is%
\begin{align}
P_{b_{22}}|_{\mathbb{B}_{323}}\mathrm{P\!}\left( \hat{b}%
_{11}^{(1)}|_{b_{11}^{(0)}\text{,}s_{2}^{(0)},s_{3}^{(2)}}\right) & =\mathrm{%
P\!}\left( \mathfrak{q}_{2}\leq\alpha_{2}A_{0\acute{1}\acute{1}}\text{, }%
\mathfrak{n}_{2}\leq\alpha_{2}A_{\acute{1}\acute{1}1}\right)  \notag \\
& =Q\left( \sqrt{\gamma_{3,5}}\right) Q\left( \sqrt{\gamma_{3,3}}\right) .
\end{align}

$\boldsymbol{Case}$ $\boldsymbol{4}$: $s_{1}^{(0)}$ $\left(
b_{11}^{(0)}\right) $, $\hat{s}_{1}^{(2)}$ $\left( \hat{b}_{11}^{(1)}\right) 
$, $s_{2}^{(0)}$, $s_{3}^{(0)}$:

The probability of error for $b_{21}$can be computed as:%
\begin{align}
P_{b_{21}}|_{\mathbb{B}_{324}}\mathrm{P\!}\left( \hat{b}%
_{11}^{(1)}|_{b_{11}^{(0)}\text{,}s_{2}^{(0)},s_{3}^{(0)}}\right) & =\mathrm{%
P\!}\left( \mathfrak{n}_{2}\leq\alpha_{2}A_{\acute{2}\acute{1}\acute{1}}%
\text{, }\mathfrak{n}_{2}\leq\alpha_{2}A_{\acute{1}\acute{1}\acute{1}}\right)
\notag \\
& =\mathrm{P\!}\left( \mathfrak{n}_{2}\leq\alpha_{2}A_{\acute{2}\acute {1}%
\acute{1}}\right) \\
& =Q\left( \sqrt{\gamma_{3,10}}\right)
\end{align}
where $\gamma_{3,10}=\alpha_{2}^{2}A_{211}^{2}/\sigma_{\mathfrak{n}_{2}}^{2}$
and $\mathbb{B}_{324}\rightarrow\left\{ b_{11}^{(0)}\text{, }\hat{b}%
_{11}^{(1)}\text{, }s_{2}^{(0)}\text{, }s_{3}^{(0)}\right\} $.

For $b_{22}$ error probability%
\begin{align}
P_{b_{22}}|_{\mathbb{B}_{324}}\mathrm{P\!}\left( \hat{b}%
_{11}^{(1)}|_{b_{11}^{(0)}\text{,}s_{2}^{(0)},s_{3}^{(0)}}\right) & =\mathrm{%
P\!}\left( \mathfrak{q}_{2}\leq\alpha_{2}A_{0\acute{1}\acute{1}}\right) 
\mathrm{P\!}\left( \mathfrak{n}_{2}\leq\alpha_{2}A_{\acute{1}\acute{1}\acute{%
1}}\right)  \notag \\
& =Q\left( \sqrt{\gamma_{3,5}}\right) Q\left( \sqrt{\gamma_{3,4}}\right) .
\end{align}
By considering the other $60$ cases, $\boldsymbol{Case}$ $\boldsymbol{5}$ to 
$\boldsymbol{Case}$ $\boldsymbol{64}$, the total BER of the second user when 
$\hat{b}_{11}\neq$ $b_{11}$\ can be represented as%
\begin{multline}
P_{U_{2}}^{(2)}=\frac{1}{4}\left[ Q\left( \sqrt{\gamma_{3,5}}\right) \left(
\sum\limits_{v=1}^{4}Q\left( \sqrt{\gamma_{3,v}}\right) \right)
+\sum\limits_{i}d_{i}Q\left( \sqrt{\gamma_{3,i}}\right) \right] \text{, } \\
i=\left[ 1,2,7,8,9,10\right] \text{, }\mathbf{d}=\left[ 1,1,-1,-1,1,1\right]
.  \label{Eb2U11}
\end{multline}
The total BER for the second user is evaluated by combining (\ref{EB2_111})
and (\ref{Eb2U11})%
\begin{equation}
P_{U_{2}}=\frac{1}{4}\sum\limits_{i=1}^{10}g_{i}Q\left( \sqrt{\gamma_{3,i}}%
\right) \text{, }\mathbf{g}=\left[ 1,1,-1,-1,6,2,-1,-1,1,1\right] .
\label{Eb2u_avg11}
\end{equation}

\subsection{BER of Third User $\left( U_{3}|_{N=3}\right) $}

The error probability of the third user is calculated based on Fig. \ref%
{U3_cases}. The BER of $U_{3}$ depends on $s_{1}$, $s_{2}$, $s_{3}$, $\hat{s}%
_{1}$ and $\hat{s}_{2}$ and . Therefore, the average BER for $b_{3i}$ is the
average of all possible combinations.%
\begin{align}
P_{b_{3i}} &
=\sum_{g,k,v,l,c}P_{b_{3i}}|_{s_{1}^{(g)},s_{2}^{(k)},s_{3}^{(v)},\hat{s}%
_{1}^{(l)},\hat{s}_{2}^{(c)}}  \notag \\
& \text{ \ \ \ \ \ \ \ \ \ \ \ \ \ \ \ \ \ \ \ \ \ \ \ \ \ \ \ \ \ \ \ }%
\times\mathrm{P\!}\left( s_{1}^{(g)},s_{2}^{(k)},s_{3}^{(v)},\hat{s}%
_{1}^{(l)},\hat{s}_{2}^{(c)}\right)  \notag \\
& =\sum_{g,k,v,l,c}P_{b_{3i}}|_{s_{1}^{(g)},s_{2}^{(k)},s_{3}^{(v)},\hat {s}%
_{1}^{(l)},\hat{s}_{2}^{(c)}}  \notag \\
& \text{ \ \ \ \ \ \ \ \ \ \ }\times\mathrm{P\!}\left( \hat{s}_{1}^{(l)},%
\hat{s}_{2}^{(c)}|_{s_{1}^{(g)},s_{2}^{(k)},s_{3}^{(v)}}\right) \mathrm{P\!}%
\left( s_{1}^{(g)},s_{2}^{(k)},s_{3}^{(v)}\right)  \notag \\
& =\frac{1}{64}\sum_{\left\{ g,k,v,l,c\right\}
=0}^{3}P_{b_{3i}}|_{s_{1}^{(g)},s_{2}^{(k)},s_{3}^{(v)},\hat{s}_{1}^{(l)},%
\hat{s}_{2}^{(c)}}  \notag \\
& \text{ \ \ \ \ \ \ \ \ \ \ \ \ \ \ \ \ \ \ \ \ \ \ \ \ \ \ \ \ \ \ \ \ \ \
\ \ }\times\mathrm{P\!}\left( \hat{s}_{1}^{(l)},\hat{s}%
_{2}^{(c)}|_{s_{1}^{(g)},s_{2}^{(k)},s_{3}^{(v)}}\right) \cdot
\label{Eb3u_1}
\end{align}%


\begin{figure*}
    \centering
    \includegraphics{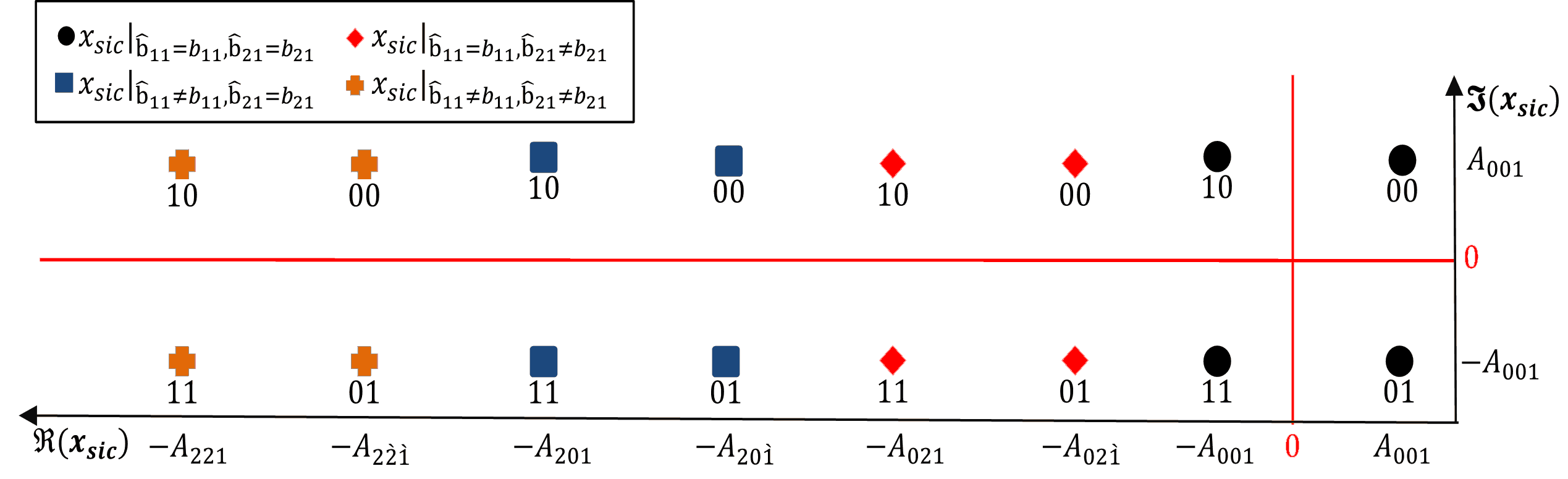}
    \caption{Equivelant constellation of the third user, $N=3$.}
    \label{U3_cases}
\end{figure*}

Table \ref{TableKey} presents the four different possible
scenarios for this user.

\begin{table}[tbp] \centering%
\caption{The four possible cases for the third user.}%
\begin{tabular}{|c|c|c|}
\hline
Scenario & $U_{1}$ Detection & $U_{2}$ Detection \\ \hline\hline
$1$ & $\checkmark$ & $\checkmark$ \\ \hline
$2$ & $\checkmark$ & $\times$ \\ \hline
$3$ & $\times$ & $\checkmark$ \\ \hline
$4$ & $\times$ & $\times$ \\ \hline
\end{tabular}
\label{TableKey}%
\end{table}
It should be noted that both $b_{12}$ and $b_{22}$ bits do not affect the
error probability of the third user, hence it is assumed that $%
P_{b_{12}}=P_{b_{22}}=1$. As for the first scenario where $\hat{b}%
_{11}=b_{11}$ and $\hat{b}_{21}=b_{21}$, the error probability of the third
user is derived according to Fig. \ref{U3_cases} (solid circles) and (\ref%
{Eb3u_1}).

$\boldsymbol{Case}$ $\boldsymbol{1}$: $s_{1}^{(0)}$ $\left(
b_{11}^{(0)}\right) $, $\hat{s}_{1}^{(0)}$ $\left( \hat{b}_{11}^{(0)}\right) 
$, $s_{2}^{(2)}$ $\left( b_{21}^{(1)}\right) $, $\hat{s}_{2}^{(2)}$ $\left( 
\hat{b}_{21}^{(1)}\right) ,$ $s_{3}^{(2)}$:

The transmitted signal $x=A_{1\acute{1}\acute{1}}+jA_{111}$ is subtracted by 
$\hat{s}_{1}$ and $\hat{s}_{2}$, hence, $x_{sic}=A_{00\acute{1}}+jA_{001}$.
The error probability for this case can be obtained as follows.%
\begin{align}
P_{b_{31}}|_{\mathbb{A}_{331}}\mathrm{P\!}\left( \hat{b}_{11}^{(0)},\hat {b}%
_{21}^{(1)}|_{\ddot{\mathbb{A}}_{331}}\right) & =\mathrm{P\!}\left( 
\mathfrak{n}_{3}\geq\alpha_{3}A_{001},\mathfrak{n}_{3}\geq\alpha_{3}A_{%
\acute{1}11}\right)  \notag \\
& =Q\left( \sqrt{\gamma_{3,11}}\right)
\end{align}
$\mathbb{A}_{331}\rightarrow\left\{ b_{11}^{(0)}\text{, }\hat{b}_{11}^{(0)}%
\text{, }b_{21}^{(1)}\text{, }\hat{b}_{21}^{(1)}\text{, }s_{3}^{(2)}\right\} 
$, $\ddot{\mathbb{A}}_{331}\rightarrow\left\{ b_{11}^{(0)}\text{, }%
b_{21}^{\left( 1\right) }\text{, }s_{3}^{(2)}\right\} $, $\mathfrak{n}%
_{3}\triangleq\mathfrak{R}\left( \check{w}_{3}\right) $ and $%
\gamma_{3,11}=\alpha_{3}^{2}A_{001}^{2}/\sigma_{\mathfrak{n}_{3}}^{2}$.

The second bit $b_{32\text{ }}$error probability is derived as follows.%
\begin{align}
P_{b_{32}}|\mathbb{A}_{331}\mathrm{P\!}\left( \hat{b}_{11}^{(0)},\hat{b}%
_{21}^{(1)}|_{\ddot{\mathbb{A}}_{331}}\right) \!\!\!\!\!\! & =\!\!\!\!\!\!%
\mathrm{P\!}\left( \mathfrak{q}_{3}\leq\alpha_{3}A_{00\acute{1}}\right) 
\mathrm{P\!}\left( \mathfrak{n}_{3}\geq\alpha_{3}A_{\acute{1}11}\right) 
\notag \\
& =\!\!\!\!Q\left( \sqrt{\gamma_{3,11}}\right) \left( 1-Q\left( \sqrt{%
\gamma_{3,1}}\right) \right) \!\!.
\end{align}
where $\mathfrak{q}_{3}\triangleq\mathfrak{I}\left( \check{w}_{3}\right) $.

$\boldsymbol{Case}$ $\boldsymbol{2}$: $s_{1}^{(0)}$ $\left(
b_{11}^{(0)}\right) $, $\hat{s}_{1}^{(0)}$ $\left( \hat{b}_{11}^{(0)}\right) 
$, $s_{2}^{(2)}$ $\left( b_{21}^{(1)}\right) $, $\hat{s}_{2}^{(2)}$ $\left( 
\hat{b}_{21}^{(1)}\right) $, $s_{3}^{(0)}$:

The error probability of this case is expressed as%
\begin{align}
P_{b_{31}}|_{\mathbb{A}_{332}}\mathrm{P\!}\left( \hat{b}_{11}^{(0)},\hat {b}%
_{21}^{(1)}|_{\ddot{\mathbb{A}}_{332}}\right) & =\mathrm{P\!}\left(
\alpha_{3}A_{\acute{1}1\acute{1}}\leq\mathfrak{n}_{3}\leq\alpha_{3}A_{00%
\acute{1}}\right)  \notag \\
& =Q\left( \sqrt{\gamma_{3,11}}\right) -Q\left( \sqrt{\gamma_{3,2}}\right) .
\end{align}
$\mathbb{A}_{332}\rightarrow\left\{ b_{11}^{(0)}\text{, }\hat{b}_{11}^{(0)}%
\text{, }b_{21}^{(1)}\text{, }\hat{b}_{21}^{(1)}\text{, }s_{3}^{(0)}\right\} 
$ and $\ddot{\mathbb{A}}_{332}\rightarrow\left\{ b_{11}^{(0)}\text{, }%
b_{21}^{(1)}\text{, }s_{3}^{(0)}\right\} $.

For $b_{32\text{ }}$, the error probability can be obtained as%
\begin{align}
P_{b_{32}}|_{\mathbb{A}_{332}}\mathrm{P\!}\left( \hat{b}_{11}^{(0)},\hat {b}%
_{21}^{(1)}|_{\ddot{\mathbb{A}}_{332}}\right) \!\! & =\!\!\mathrm{P\!}\left( 
\mathfrak{q}_{3}\leq\alpha_{3}A_{00\acute{1}}\right) \mathrm{P\!}\left( 
\mathfrak{n}_{3}\geq\alpha_{3}A_{\acute{1}1\acute{1}}\right)  \notag \\
& =Q\left( \sqrt{\gamma_{3,11}}\right) \left( 1-Q\left( \sqrt {\gamma_{3,2}}%
\right) \right) .
\end{align}

$\boldsymbol{Case}$ $\boldsymbol{3}$: $s_{1}^{(0)}$ $\left(
b_{11}^{(0)}\right) $, $\hat{s}_{1}^{(2)}$ $\left( \hat{b}_{11}^{(0)}\right) 
$, $s_{2}^{(0)}$ $\left( b_{21}^{(0)}\right) $, $\hat{s}_{2}^{(0)}$ $\left( 
\hat{b}_{21}^{(0)}\right) $, $s_{3}^{(2)}$:

The error probability of this case is derived as%
\begin{align}
P_{b_{32}}|_{\mathbb{A}_{333}}\mathrm{P\!}\left( \hat{b}_{11}^{(0)},\hat {b}%
_{21}^{(0)}|_{\ddot{\mathbb{A}}_{333}}\right) & =\mathrm{P\!}\left( 
\mathfrak{n}_{3}\geq\alpha_{3}A_{001}\right)  \notag \\
& =Q\left( \sqrt{\gamma_{3,11}}\right) .
\end{align}

$\mathbb{A}_{333}\rightarrow\left\{ b_{11}^{(0)},\hat{b}%
_{11}^{(0)},b_{21}^{(0)},\hat{b}_{21}^{(0)},s_{3}^{(2)}\right\} $ and $\ddot{%
\mathbb{A}}_{333}\rightarrow\left\{
b_{11}^{(0)},b_{21}^{(0)},s_{3}^{(2)}\right\} $.

The error probability for $b_{32}$ is 
\begin{align}
P_{b_{32}}|_{\mathbb{A}_{333}}\mathrm{P\!}\left( \hat{b}_{11}^{(0)},\hat {b}%
_{21}^{(0)}|_{\ddot{\mathbb{A}}_{333}}\right) \!\! & =\!\!\mathrm{P\!}\left( 
\mathfrak{q}_{3}\leq\alpha_{3}A_{00\acute{1}}\right) \mathrm{P\!}\left( 
\mathfrak{n}_{3}\geq\alpha_{3}A_{\acute{1}\acute{1}1}\right)  \notag \\
& =Q\left( \sqrt{\gamma_{3,11}}\right) \left( 1-Q\left( \sqrt {\gamma_{3,3}}%
\right) \right) .
\end{align}

$\boldsymbol{Case}$ $\boldsymbol{4}$: $s_{1}^{(0)}$ $\left(
b_{11}^{(0)}\right) $, $\hat{s}_{1}^{(2)}$ $\left( \hat{b}_{11}^{(0)}\right) 
$, $s_{2}^{(0)}$ $\left( b_{21}^{(0)}\right) $, $\hat{s}_{2}^{(0)}$ $\left( 
\hat{b}_{21}^{(0)}\right) $, $s_{3}^{(0)}$:

The error probability of this case is obtained as%
\begin{align}
P_{b_{31}}|_{\mathbb{A}_{334}}\mathrm{P\!}\left( \hat{b}_{11}^{(0)},\hat {b}%
_{21}^{(0)}|_{\ddot{\mathbb{A}}_{334}}\right) & =\mathrm{P\!}\left( 
\mathfrak{n}_{3}\leq\alpha_{3}A_{00\acute{1}}\right)  \notag \\
& \text{ \ \ \ \ \ \ \ \ \ \ \ }-\mathrm{P\!}\left( \mathfrak{n}%
_{3}\leq\alpha_{3}A_{\acute{1}\acute{1}\acute{1}}\right)  \notag \\
& =Q\left( \sqrt{\gamma_{3,11}}\right) -Q\left( \sqrt{\gamma_{3,4}}\right) .
\end{align}

$\mathbb{A}_{334}\rightarrow\left\{ s_{1}^{(0)},\hat{s}%
_{1}^{(2)},s_{2}^{(0)},\hat{s}_{2}^{(0)},s_{3}^{(0)}\right\} $ and $\ddot{%
\mathbb{A}}_{334}\rightarrow\left\{
b_{11}^{(0)},b_{21}^{(0)},s_{3}^{(0)}\right\} $.

For $b_{32}$ is as follows%
\begin{align}
P_{b_{32}}|\mathbb{A}_{334}\mathrm{P\!}\left( \hat{b}_{11}^{(0)},\hat{b}%
_{21}^{(0)}|_{\ddot{\mathbb{A}}_{334}}\right) & =\mathrm{P\!}\left( 
\mathfrak{q}_{3}\leq\alpha_{3}A_{00\acute{1}}\right) \mathrm{P\!}\left( 
\mathfrak{n}_{3}\geq\alpha_{3}A_{\acute{1}\acute{1}\acute{1}}\right)  \notag
\\
& =Q\left( \sqrt{\gamma_{3,11}}\right) \left( 1-Q\left( \sqrt {\gamma_{3,4}}%
\right) \right) .
\end{align}
By taking into consideration the other $60$ cases, $\boldsymbol{Case}$ $%
\boldsymbol{5}$ to $\boldsymbol{Case}$ $\boldsymbol{64}$, which are obtained
in a similar manner, the total BER for the first scenario where $\hat{b}%
_{11}=b_{11}$ and $\hat{b}_{21}=b_{21}$ can be represented as%
\begin{multline}
P_{U_{3}}^{(1)}=\frac{1}{4}Q\left( \sqrt{\gamma_{3,11}}\right) \left(
8-\sum_{i=1}^{4}Q\left( \sqrt{\gamma_{3,i}}\right) \right)  \label{Eb31} \\
-\frac{1}{4}\left[ Q\left( \sqrt{\gamma_{3,2}}\right) -Q\left( \sqrt{%
\gamma_{3,4}}\right) \right] .
\end{multline}

As for the other scenarios, a similar procedure is followed as for the case $%
\hat{b}_{11}=b_{11}$ and $\hat{b}_{21}=b_{21}$. The analysis for $\hat {b}%
_{11}=b_{11}$, $\hat{b}_{21}\neq b_{21}$, $\hat{b}_{11}\neq b_{11}$, $\hat{b}%
_{21}=b_{21}$ and $\hat{b}_{11}\neq b_{11},$ $\hat{b}_{21}\neq b_{21}$ is
based on Fig. \ref{U3_cases} and (\ref{Eb3u_1}). The final total BER results
for this scenario is respectively given by, 
\begin{multline}
P_{U_{3}}^{(2)}=2-Q\left( \sqrt{\gamma_{3,1}}\right) -Q\left( \sqrt {%
\gamma_{3,2}}\right) -Q\left( \sqrt{\gamma_{3,4}}\right) -Q\left( \sqrt{%
\gamma_{3,14}}\right) \\
+2Q\left( \sqrt{\gamma_{3,13}}\right) +Q\left( \sqrt{\gamma_{3,11}}\right)
\left( 4-\sum_{i=1}^{4}Q\left( \sqrt{\gamma_{3,i}}\right) \right)
\label{Eb3_1}
\end{multline}%
\begin{equation}
P_{U_{3}}^{(3)}=Q\left( \sqrt{\gamma_{3,11}}\right) \left(
\sum_{i=1}^{4}Q\left( \sqrt{\gamma_{3,i}}\right) \right) +Q\left( \sqrt{%
\gamma _{3,2}}\right) +Q\left( \sqrt{\gamma_{3,4}}\right)  \label{Eb3_2}
\end{equation}
and%
\begin{multline}
P_{U_{3}}^{(4)}=Q\left( \sqrt{\gamma_{3,1}}\right) -Q\left( \sqrt {%
\gamma_{3,15}}\right) +\sum_{i=16}^{18}Q\left( \sqrt{\gamma_{3,i}}\right)
\label{Eb3_3} \\
+Q\left( \sqrt{\gamma_{3,11}}\right) \left( \sum_{i=1}^{4}Q\left( \sqrt{%
\gamma_{3,i}}\right) \right)
\end{multline}
where $\gamma_{3,12}=\alpha_{3}^{2}A_{02\acute{1}}^{2}/\sigma_{\mathfrak{n}%
_{3}}^{2}$, $\gamma_{3,13}=\alpha_{3}^{2}A_{201}^{2}/\sigma_{\mathfrak{n}%
_{3}}^{2}$, $\gamma_{3,14}=\alpha_{3}^{2}A_{20\acute{1}}^{2}/\sigma_{%
\mathfrak{n}_{3}}^{2}$, $\gamma_{3,15}=\alpha_{3}^{2}A_{2\grave{2}\acute{1}%
}^{2}/\sigma_{\mathfrak{n}_{3}}^{2}$, $\gamma_{3,16}=\alpha_{3}^{2}A_{2%
\grave{2}1}^{2}/\sigma_{\mathfrak{n}_{3}}^{2}$, $\gamma_{3,17}=%
\alpha_{3}^{2}A_{22\acute{1}}^{2}/\sigma_{\mathfrak{n}_{3}}^{2}$, $%
\gamma_{3,18}=\alpha_{3}^{2}A_{221}^{2}/\sigma_{\mathfrak{n}_{3}}^{2}$. By
combining the results in (\ref{Eb31}), (\ref{Eb3_1}), (\ref{Eb3_2}) and (\ref%
{Eb3_3}), the total BER for the third user can be computed as 
\begin{multline}
P_{U_{3}}=\frac{1}{4}\left[ \left( \sum_{i}v_{i}Q\left( \sqrt{\gamma_{3,i}}%
\right) \right) \right] \text{,} \\
\text{ }i=\left[ 2,4,11,13,14,15,16,17,18\right] \text{,} \\
\mathbf{v}=[-1,-1,12,2,-1,-1,-1,-1,-1].
\end{multline}

\subsection{Average BER, $N=3$\label{SubSec-averageBER U3}}

Similar to the $N=2$ NOMA system, the average BER of $N=3$ NOMA system can
be evaluated by averaging over the PDFs of all $\gamma_{n,c}$ values, which
are given in Appendix I. Therefore, by substituting $N=3$ and $n=\left[ 1,2,3%
\right] $ in the ordered PDF in \ (\ref{E-Ordered-PDF}), the exact average
BER of the first and second users can be simplified to 
\begin{multline}
\bar{P}_{U_{1}}=\frac{3}{4\pi\Gamma(m)}\sum\limits_{c=1}^{4}\sum_{k=0}^{2}%
\sum_{i=0}^{\infty}\binom{2}{k}\left( -1\right) ^{k}S_{i}\left( \frac {m}{%
\overline{\gamma}_{1,c}}\right) ^{m(1+k)}  \label{NakagamiU1_3U} \\
\times\int_{0}^{\frac{\pi}{2}}\frac{\left( i+mk\right) !}{\left( \frac {1}{%
2\sin^{2}\left( \psi_{1,c}\right) }+\frac{m\text{ }\left( 1+k\right) }{%
\overline{\gamma}_{1,c}}\right) ^{i+mk+1}}d\psi_{1,c}
\end{multline}%
\begin{multline}
\bar{P}_{U_{2}}=\frac{3}{2\pi\Gamma(m)}\sum\limits_{c=1}^{10}\sum_{k=0}^{1}%
\sum_{i=0}^{\infty}\left( -1\right) ^{k}S_{i}g_{c}\left( \frac {m}{\overline{%
\gamma}_{2,c}}\right) ^{m(2+k)} \\
\times\int_{0}^{\frac{\pi}{2}}\frac{\left( i+m(1+k)\right) !}{\left( \frac{1%
}{2\sin^{2}\left( \psi_{2,c}\right) }+\frac{m\text{ }\left( 2+k\right) }{%
\overline{\gamma}_{2,c}}\right) ^{i+m(1+k)+1}}d\psi _{2,c}\text{,} \\
\mathbf{g}=\left[ 1,1,-1,-1,6,2,-1,-1,1,1\right]  \label{NakagamiU2_3U}
\end{multline}
and%
\begin{multline}
\bar{P}_{U_{3}}=\frac{3}{8\pi\Gamma(m)}\sum\limits_{c=1}^{18}\sum
_{i=0}^{\infty}v_{c}S_{i}\left( \frac{m}{\overline{\gamma}_{3,c}}\right)
^{3m} \\
\times\int_{0}^{\frac{\pi}{2}}\frac{\left( i+2m\right) !}{\left( \frac {1}{%
2\sin^{2}\left( \psi_{3,c}\right) }+\frac{3m\text{ }}{\overline{\gamma }%
_{3,c}}\right) ^{i+2m+1}}d\psi_{3,c}\text{,} \\
c\in\left\{ 2,4,11,13,14,15,16,17,18\right\} \text{,}  \label{Nakagami3U_3U}
\\
\mathbf{v}=[-1,-1,12,2,-1,-1,-1,-1,-1].
\end{multline}
The closed-form average BER for the first, second, and third users over\
Rayleigh fading channel $\left( m=1\right) $ are shown in (\ref{Ray_3U_1U1}%
), (\ref{Ray_3U_2U}), and (\ref{Ray_3U_3U22}), respectively.%
\begin{equation}
\bar{P}_{U_{1}}=\frac{1}{4}\dsum \limits_{c=1}^{4}\left( 1-\sqrt{\frac{2%
\overline{\gamma}_{_{3,c}}}{2\overline{\gamma}_{3,c}+3}}\right)
\label{Ray_3U_1U1}
\end{equation}%
\begin{multline}
\bar{P}_{U_{2}}=\frac{1}{4}\left[ \dsum \limits_{c=1}^{10}g_{c}\left( \sqrt{%
\frac{2\overline{\gamma}_{3,c}}{2\overline{\gamma}_{3,c}+3}}-\frac{3}{2}%
\sqrt{\frac{\overline{\gamma}_{3,c}}{\overline{\gamma}_{3,c}+1}}+\frac{1}{2}%
\right) \right] \text{,} \\
\mathbf{g}=\left[ 1,1,-1,-1,6,2,-1,-1,1,1\right]  \label{Ray_3U_2U}
\end{multline}
and%
\begin{multline}
\bar{P}_{U_{3}}=\frac{1}{4}\sum_{c}v_{c}\left( -\sqrt{\frac{2\overline {%
\gamma}_{3,c}}{2\overline{\gamma}_{3,c}+3}}-3\sqrt{\frac{2\overline{\gamma }%
_{3,c}}{2\overline{\gamma}_{3,c}+1}}\right. \\
+\left. 3\sqrt{\frac{\overline{\gamma}_{3,c}}{\overline{\gamma}_{3,c}+1}}+%
\frac{1}{2}\right) \\
c\in\left\{ 2,4,11,13,14,15,16,17,18\right\} , \\
\mathbf{v}=[-1,-1,12,2,-1,-1,-1,-1,-1].  \label{Ray_3U_3U22}
\end{multline}

\section{\label{Optimality}Optimum Power Allocation Problem}

The optimal power allocation problem is formulated for two different
objective functions using the derived exact BER expressions for $N=2$ and $%
N=3$ NOMA\ systems. The first objective is to obtain the optimum power
coefficients $\ddot{\beta}_{n}$ that minimize the overall average BER of all
users. The second objective function is to evaluate the optimum power
coefficients which guarantee fairness among all users. The fairness in this
work indicates equal BER for all the users.

The optimal power allocation for minimizing the average BER is formulated as
follows: 
\begin{subequations}
\begin{equation}
\min_{\beta_{n}}\text{ }\frac{1}{N}\sum_{n=1}^{N}\bar{P}_{U_{n}}
\end{equation}
subject to:%
\begin{gather}
\sum_{n=1}^{N}\beta_{n}=1  \label{OptimalAVERAEGE} \\
\beta_{l}\geqslant\beta_{k}\text{, }l\neq k\text{, }l<k\text{, }\left\{
l,k\right\} \in\left\{ 1,2,\ldots,N\right\}
\end{gather}
where the first constraint limits the maximum transmit power, which is
normalized to unity. The second constraint is used assure that the power
allocated to each user is inversely proportional to its channel gain, i.e., $%
\beta_{1}>\beta_{2}>\cdots>\beta_{N}$ are assigned for the users with the
channel gains $\alpha_{1}<\alpha_{2}<\cdots<\alpha_{N}$, respectively. The
problem in (\ref{OptimalAVERAEGE}) is a constrained non-linear optimization
problem which is solved using the Interior-Point Optimization (IPO)
algorithm \cite{0806023}.

The optimal power allocation for achieving fairness among the users in the
NOMA system is formulated as follows: 
\end{subequations}
\begin{equation}
\bar{P}_{U_{l}}=\bar{P}_{U_{k}}\text{, }\forall\{l,k\}\in\left\{
1,2,\ldots,N\right\} \text{, }l\neq k  \label{Optimalfair}
\end{equation}
The constraints for the second optimization problem are similar to those in
the first optimization problem and the solution can be obtained using the
same approach.

\section{\label{Results}Numerical and Simulation Results}

This section presents numerical and Monte Carlo simulation results for $N=2$
and $N=3$ downlink NOMA systems. All users are assumed to be equipped with a
single antenna, and the channel between the BS and each user is modeled as
an ordered Nakagami-$m$ flat fading channel. The randomly generated channels
are ordered based on their strength, where the weakest channel is assigned
to the first user and the strongest channel is assigned to $N$th user. The
transmitted symbols for all users are selected uniformly from a Gray coded
QPSK constellation. The total transmit power from the BS is unified for all
cases, $P_{T}=1$.


\begin{figure}[tpb]
    \centering
    \includegraphics[width=3.45in]{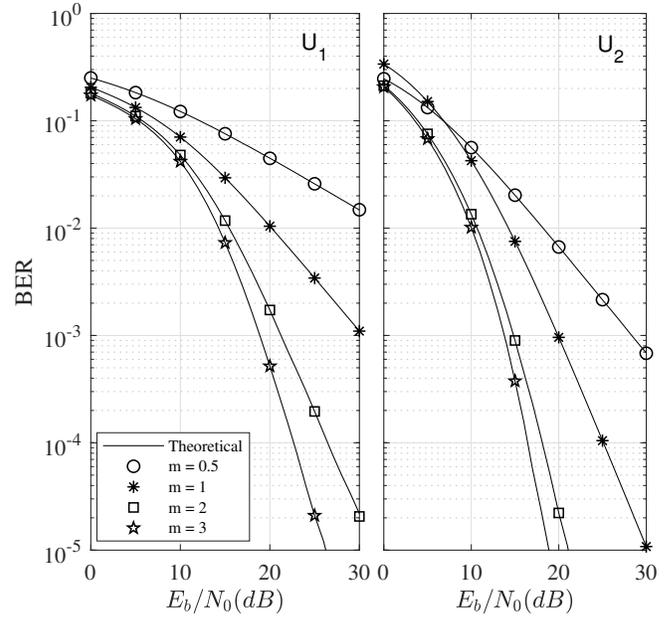}
    \caption{BER for the first and second users, $N=2$, $m=0.5,1,2$ and $3$, and $\Omega=1.$}
    \label{BER_2U}
 \end{figure}

Fig. \ref{BER_2U} presents the analytical and simulated BER performance of
the two users scenario, $N=2$ for power coefficients $\beta_{1}=0.7$ and $%
\beta_{2}=0.3$, $\Omega=1$ and various values of $m$ over a range of $%
E_{b}/N_{0}$, where $E_{b}/N_{0}=1/N_{0}$. As can be noted from the figure,
the analytical results obtained using (\ref{Naka_2U_1}), (\ref{Naka_2U_2}), (%
\ref{Pe11}), and (\ref{Pe22}) perfectly match the simulation results for all
the considered values of $m$ and $E_{b}/N_{0}$. It is worth noting that the $%
m=1$ case corresponds to the derived to the Rayleigh fading case.


\begin{figure*} 
\centering
\includegraphics[width=6.5in]{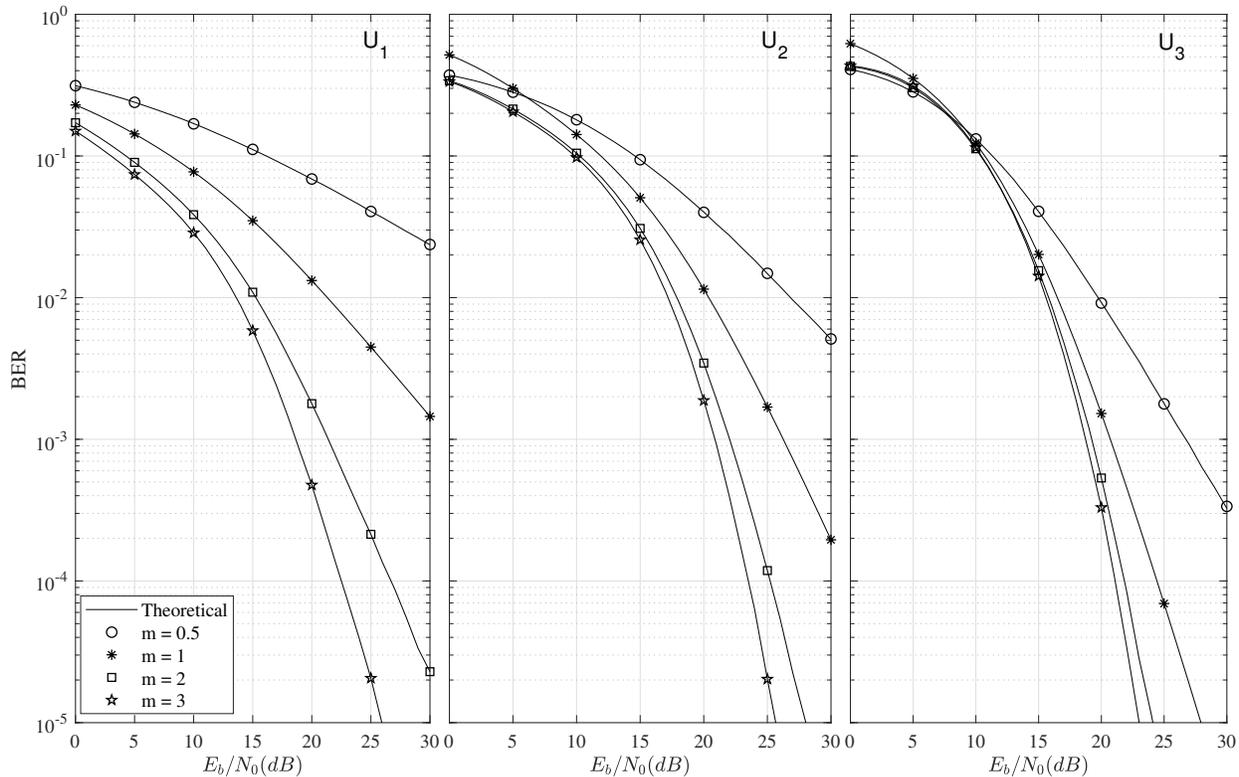}
\caption{BER for the first, second, and third users in the $N=2$, $m=0.5,1,2 $ and $3$, and $\Omega = 1.$}
\label{BER_3U}
\end{figure*}

Fig. \ref{BER_3U} is generally similar to Fig. \ref{BER_2U}, except that it
considers the three users scenario, i.e., $N=3$. The power allocation
coefficients are $\beta_{1}=0.8$, $\beta_{2}=0.15$ and $\beta_{3}=0.05$. The
figure clearly shows the perfect match between the analytical results
obtained using (\ref{NakagamiU1_3U})-(\ref{Ray_3U_3U22}) and simulation
results for all $m$ values and over the entire $E_{b}/N_{0}$ range. As can
be seen from Figures \ref{BER_2U} and \ref{BER_3U}, the performance of the
first user is more sensitive to the variations of $m$ as compared to the
second and third users, which is due to the fact that the fading effect
becomes less significant for the near users. 


\begin{figure}[tb]
    \centering
    \includegraphics[width=3.5in]{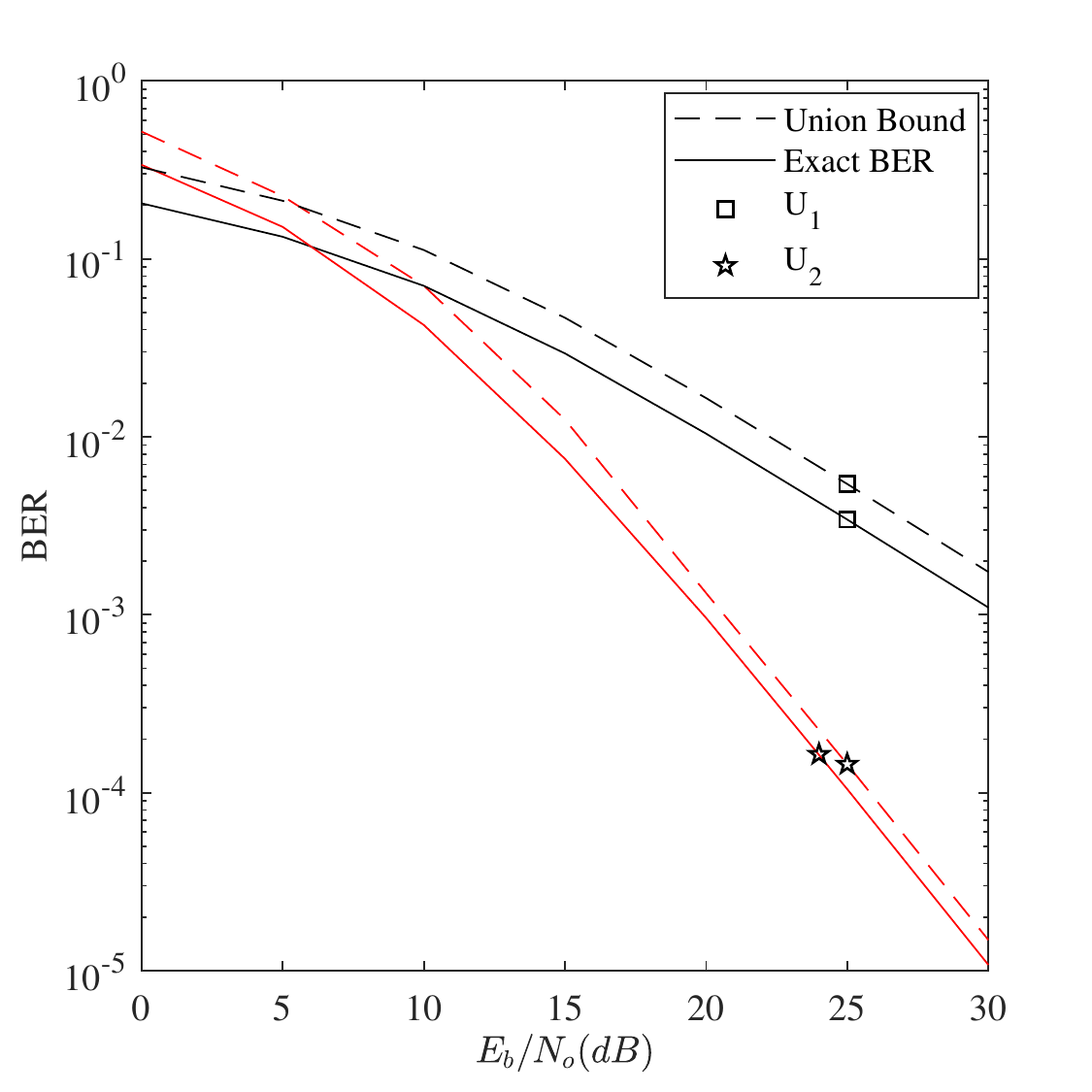}
    \caption{Exact BER and union bound for the first and second users, $N=2$, $m=1$ and $\Omega=1.$}
    \label{Unionbound}
 \end{figure}

Fig. \ref{Unionbound} compares the exact BER and union bound \cite{8501953}
for $N=2$, $m=1$, $\beta_{1}=0.7$ and $\beta_{2}=0.3$. It can be noted that
the union bound is generally tight in the high $E_{b}/N_{0}$ range,
particularly for the second user. For example, the gap between the exact BER
and union bound at $E_{b}/N_{0}=20$ dB is about $2$ dB for the first user
and $1$ dB for the second user. At low $E_{b}/N_{0}$, the gap may increase
to $3$ dB. Therefore, using the exact BER expression is critical when
accurate BER estimates are desired.


\begin{figure*}[tbp]
\centering
\includegraphics[width=6.4in]{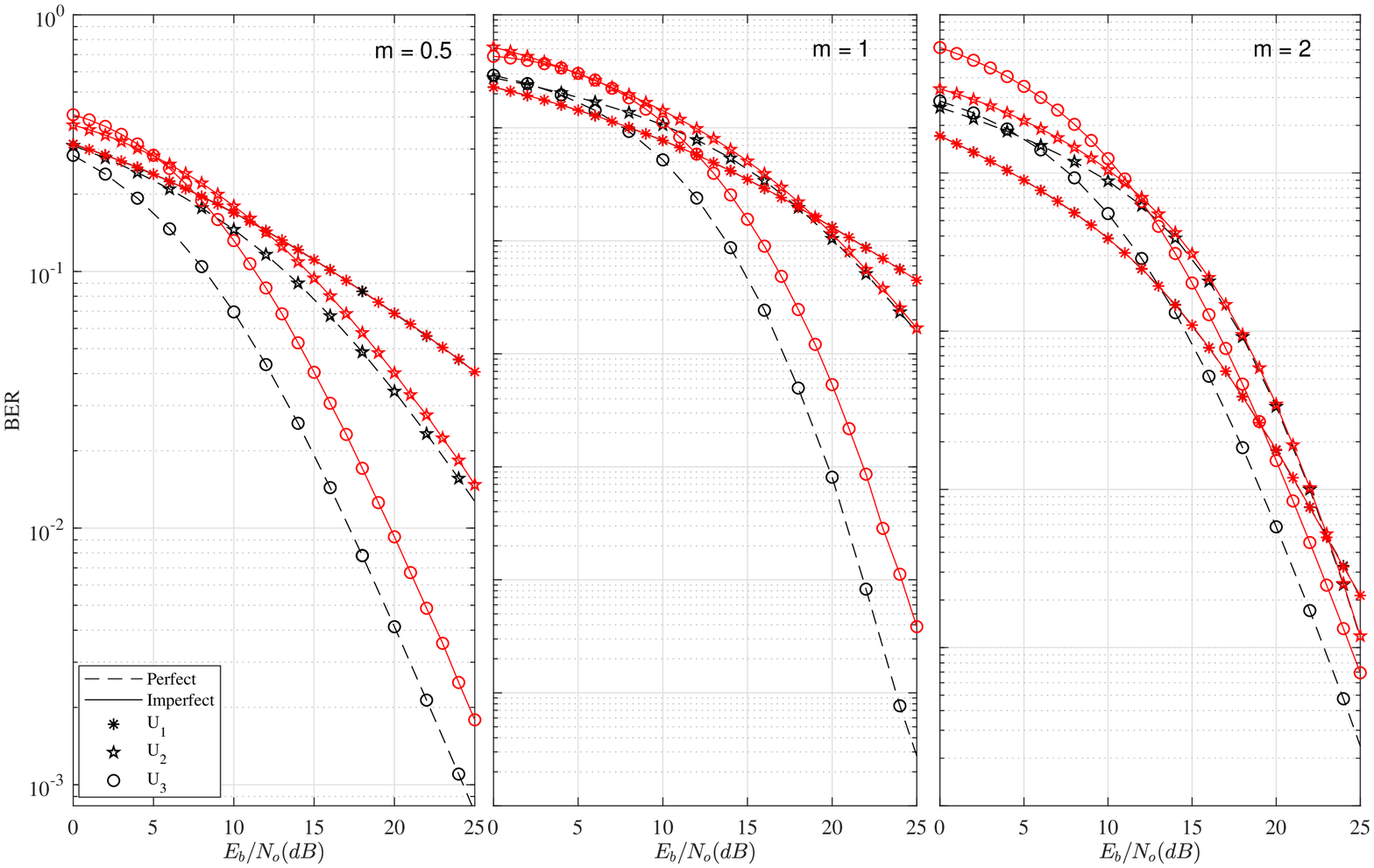}
\caption{Perfect and
imperfect BER for the first, second, and third users in the N=3, $ m= 0.5, 1$ and $2$, and $\Omega = 1$.}
\label{Imperfefct}
\end{figure*}

Fig. \ref{Imperfefct} shows the BER for $N=3$ under perfect and imperfect
SIC for different values of $m$. Although the perfect SIC assumption may
tremendously reduce the BER analysis, the results presented in Fig. \ref%
{Imperfefct} show that such assumption is too optimistic, particularly for $%
U_{3}$. As expected, the results for $U_{1}$ with/without SIC are identical
because $U_{1}$ detection does not involve a SIC process. For $U_{2}$, the
impact of the perfect SIC assumption is apparent at low SNRs, because at
high $E_{b}/N_{0}$, $U_{1}$ signal is mostly detected correctly, and thus,
the results with/without SIC converge. The third user $U_{3}$ is the one who
will experience the maximum difference because the probability that the two
SIC operations are performed successfully is relatively small. Therefore,
the BER with and without SIC for $U_{3}$ will exhibit substantial
difference. Moreover, as $m$ increases, the BER with perfect and imperfect
BERs become closer, particularly at high SNRs, which is due to that fact
that for high values of $m$ the fading is less severe, which implies that
the probability of having successful SIC operations is higher as compared to
the low $m$ values.

Tables \ref{TableKey copy(2)} and \ref{TableKey copy(4)} present the optimal
power coefficients that provide equal BER for all users. The results are
obtained for different values of $E_{b}/N_{0}$, $N=2$, $3$, and $m=1$, $3$.
As can be noted from the results in Table \ref{TableKey copy(2)}, most of
the power is actually allocated for $U_{1}$, particularly at high SNRs. For $%
N=2$, the first user is allocated more than $98\%$ of the total power at $%
E_{b}/N_{0}=30$ dB, and it is about $89\%$ for $N=3$. Moreover, the range of
values that the power coefficients might be allocated depends drastically on 
$N$. The same trends can be noted for the $m=3$ case in Table \ref{TableKey
copy(4)}. Nevertheless, the power given to the first user generally
decreases by increasing $m$ as the impact of the AWGN becomes more
noticeable.

\begin{table}[tbp] \centering%
\caption{Optimal power allocation to achieve fairness, for $m=1$.}%
\begin{tabular}{|l||l|l||l|l|l|}
\hline
& \multicolumn{2}{|c||}{$N=2$} & \multicolumn{3}{|c|}{$N=3$} \\ \hline
$E_{b}/N_{0}$ & $\beta_{1}$ & $\beta_{2}$ & $\beta_{1}$ & $\beta_{2}$ & $%
\beta_{3}$ \\ \hline\hline
$0$ & $0.838$ & $0.186$ & $0.500$ & $0.27$ & $0.23$ \\ \hline
$10$ & $0.851$ & $0.151$ & $0.790$ & $0.114$ & $0.095$ \\ \hline
$20$ & $0.916$ & $0.083$ & $0.818$ & $0.151$ & $0.029$ \\ \hline
$30$ & $0.981$ & $0.018$ & $0.890$ & $0.095$ & $0.014$ \\ \hline
\end{tabular}
\label{TableKey copy(2)}%
\end{table}%

\begin{table}[tbp] \centering%
\caption{Optimal power allocation to achieve fairness, for $m=3$.}%
\begin{tabular}{|l||l|l||l|l|l|}
\hline
& \multicolumn{2}{|c||}{$N=2$} & \multicolumn{3}{|c|}{$N=3$} \\ \hline
$E_{b}/N_{0}$ & $\beta_{1}$ & $\beta_{2}$ & $\beta_{1}$ & $\beta_{2}$ & $%
\beta_{3}$ \\ \hline\hline
$0$ & $0.830$ & $0.17$ & $0.490$ & $0.24$ & $0.27$ \\ \hline
$10$ & $0.841$ & $0.159$ & $0.700$ & $0.201$ & $0.099$ \\ \hline
$20$ & $0.903$ & $0.097$ & $0.806$ & $0.145$ & $0.049$ \\ \hline
$30$ & $0.962$ & $0.038$ & $0.850$ & $0.088$ & $0.062$ \\ \hline
\end{tabular}
\label{TableKey copy(4)}%
\end{table}%

Table \ref{TableKey copy(1)} shows the optimal power coefficients which
minimize the average BER for $N=2$, $3$, and $m=1$. As can be noted from the
table, the power allocation should be performed meticulously to achieve the
desire results. Although the exact results are different, the observations
about the power coefficients generally follows those of the BER fairness
case. Moreover, the optimum power allocation requires the knowledge of \ the 
$E_{b}/N_{0}$\ at the transmitter. For example, the value of $\beta_{2}$
drops by about $50\%$ when the $E_{b}/N_{0}$ is increased from $20$ to $30$
dB.

\begin{table}[tbp] \centering%
\caption{Optimal power allocation to achieve minimum average BER, $m=1$.}%
\begin{tabular}{|l||l|l||l|l|l|}
\hline
& \multicolumn{2}{|c||}{$N=2$} & \multicolumn{3}{|c|}{$N=3$} \\ \hline
$E_{b}/N_{0}$ & $\beta_{1}$ & $\beta_{2}$ & $\beta_{1}$ & $\beta_{2}$ & $%
\beta_{3}$ \\ \hline\hline
$0$ & $0.810$ & $0.189$ & $0.546$ & $0.320$ & $0.132$ \\ \hline
$10$ & $0.842$ & $0.157$ & $0.670$ & $0.273$ & $0.057$ \\ \hline
$20$ & $0.896$ & $0.103$ & $0.860$ & $0.116$ & $0.023$ \\ \hline
$30$ & $0.943$ & $0.056$ & $0.946$ & $0.046$ & $0.007$ \\ \hline
\end{tabular}
\label{TableKey copy(1)}%
\end{table}%


\begin{figure}
    \centering
    \includegraphics[width=3.45in]{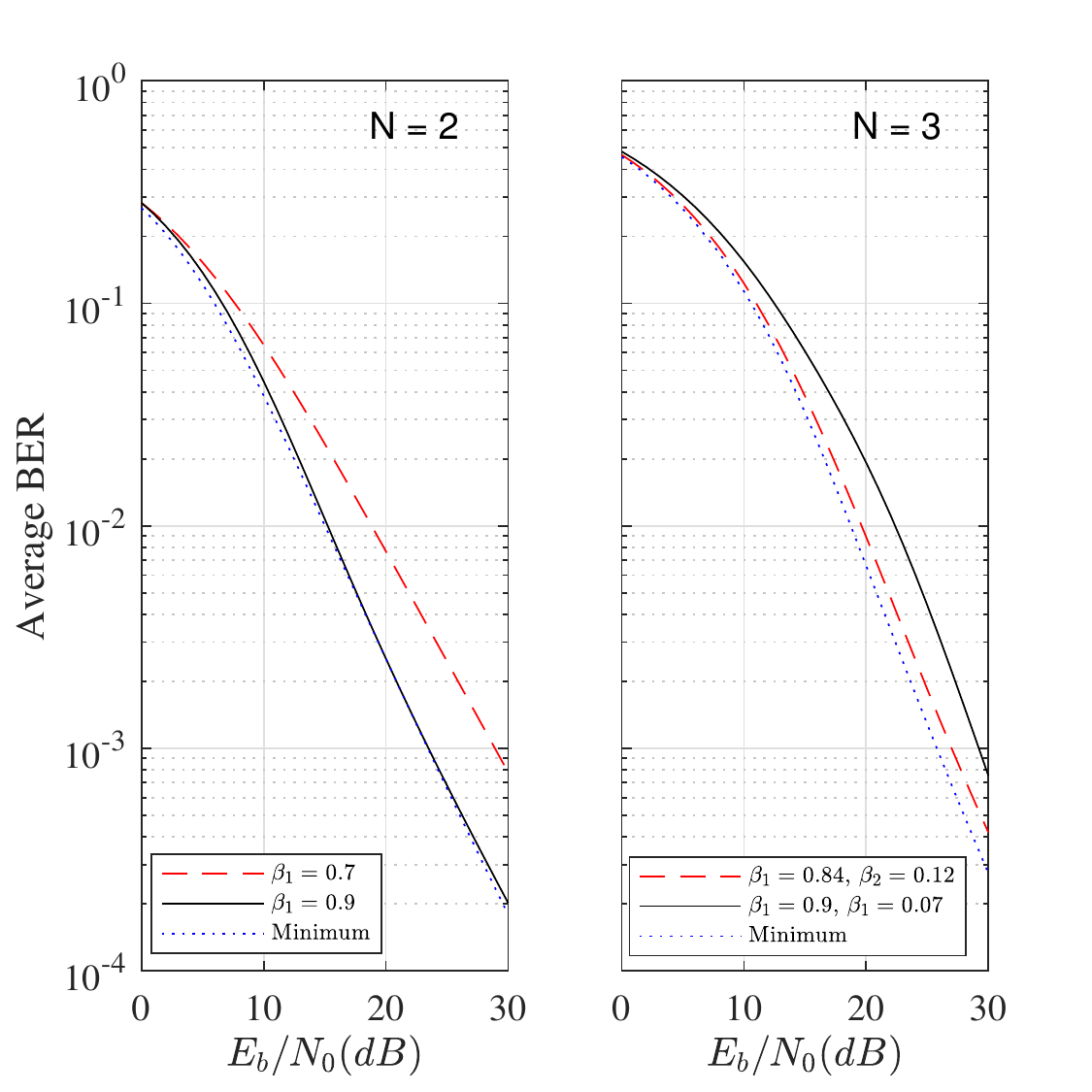}
    \caption{The average BER for $N=2$ and $N=3$ for two different power levels and the minimum value, $m=1.$}
    \label{Minaveragewithcases}
\end{figure}

Fig. \ref{Minaveragewithcases} presents the BER using the optimum power
coefficients that guarantees minimizing the average BER, in addition, two
other curves obtained using fixed power values are used where $\beta_{1}=0.7 
$, and $0.9$ for $N=2$, and $\left\{ \beta_{1}=0.84\text{, }\beta
_{2}=0.12\right\} $, and $\left\{ \beta_{1}=0.9,\beta_{2}=0.07\right\} $ for 
$N=3$. As can be noted from the figure, allocating the power coefficients
appropriately might save the need to use adaptive power values. On the other
hand, large deviations from the optimum power values might result in severe
BER degradation.


\begin{figure}
    \centering
    \includegraphics[width=3.45in]{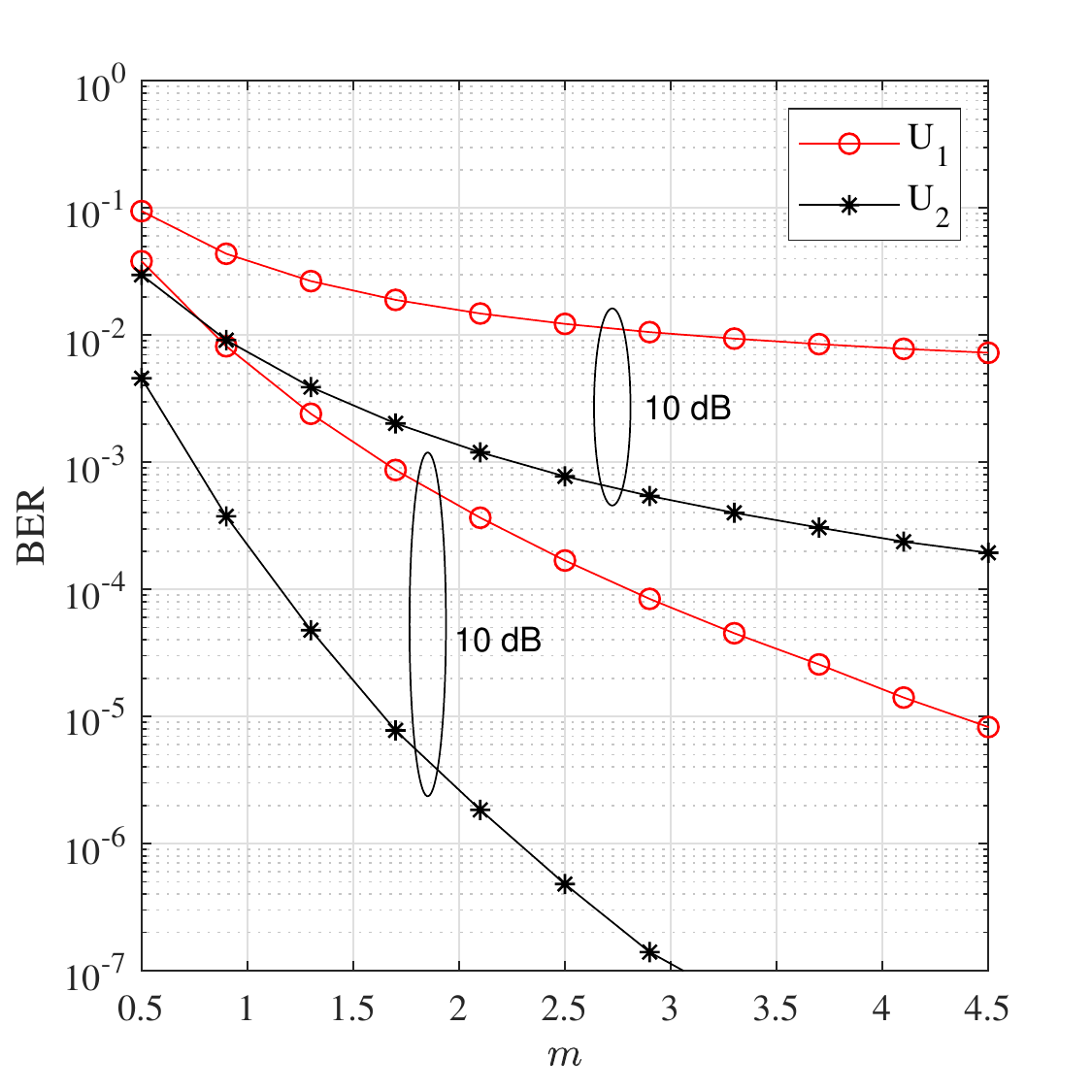}
    \caption{BER for the
first, and second users at various $m$ values, $\Omega=1$, $N=2$ NOMA system.}
    \label{fadingM2}
\end{figure}


\begin{figure}
    \centering
    \includegraphics[width=3.45in]{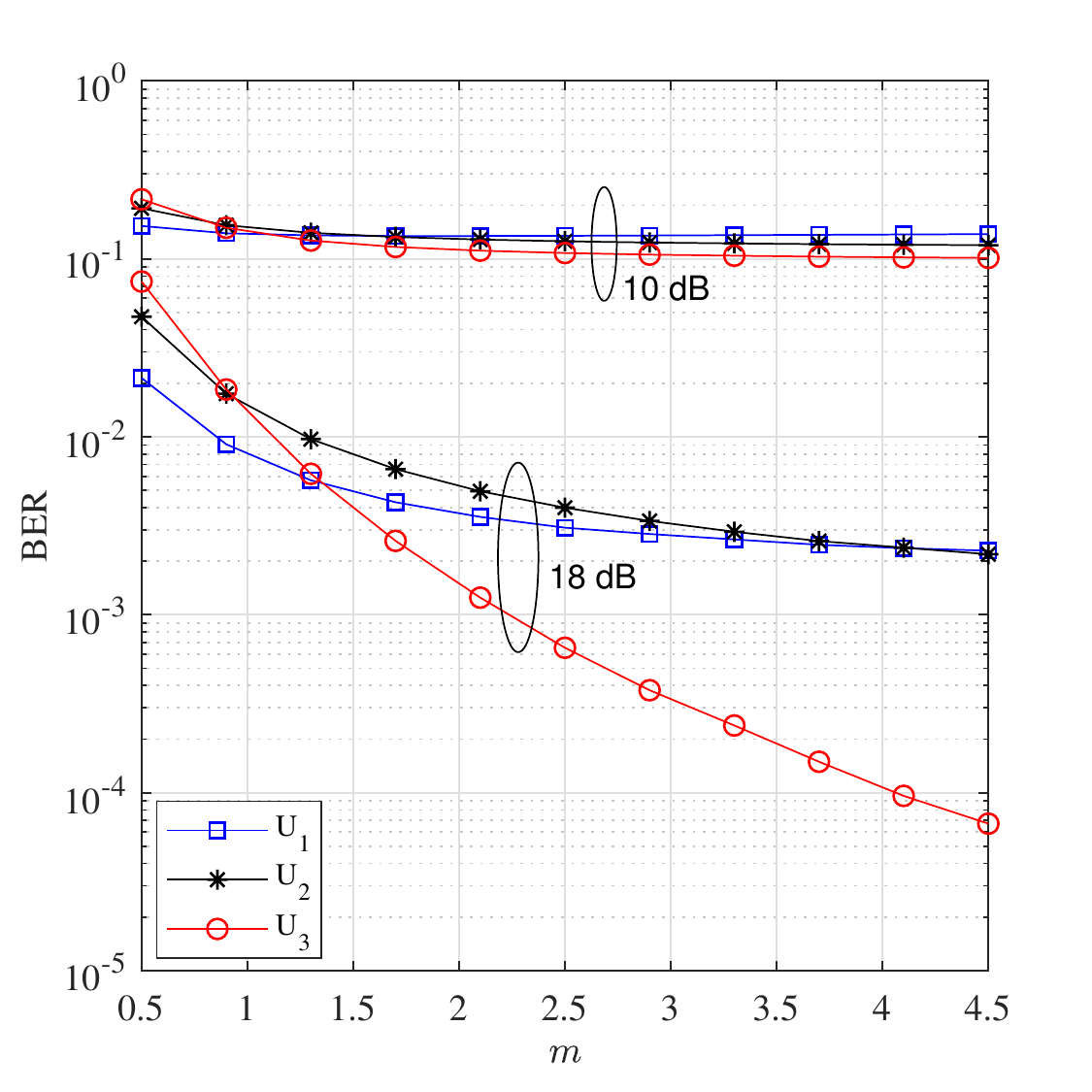}
    \caption{BER for the first,
second, and third users at various $m$ values, $\Omega=1$ and $N=3$.}
    \label{fadingM3}
\end{figure}

Figures \ref{fadingM2} and \ref{fadingM3} show the effect of the fading
factor on the performance of each user for NOMA systems with $N=2$, $3 $
where $E_{b}/N_{0}=10,$ and $18 $ dB. Power allocation coefficients at $%
E_{b}/N_{0}=10$ dB are assigned as follows, $\beta_{1}=0.84,$ $%
\beta_{2}=0.16 $ and $\beta_{1}=0.67,$ $\beta_{2}=0.27,$ and $\beta_{3}=0.06$
for the two and three users' systems, respectively. For the case where $%
E_{b}/N_{0}=18$ dB, the power allocation coefficients are $\beta_{1}=0.88,$ $%
\beta_{2}=0.12$ and $\beta_{1}=0.84,$ $\beta_{2}=0.13,$ and $\beta_{3}=0.03$
for the two and three users' systems, respectively. It should be noted that
the power allocation coefficients are the optimal values that minimize the
average BER. As can be seen from Figures \ref{fadingM2} and \ \ref{fadingM3}%
, the BER performance of all users highly depends on the fading parameter $m$%
. Additionally, it is shown that $m$ affects the performance of the higher
order users more than the lower order users, which is due to the ordering of
users based on the channel conditions that resulted in an enhanced
performance for higher order users. Moreover, when $E_{b}/N_{0}$ increases,
the effect of $m$ on the performance increases because the BER will be
mostly determined by \ the fading.

\section{\label{Conclusion}Conclusion}

This work presented the performance of a downlink NOMA system in terms of
BER where exact BER expressions were derived for different users over
Nakagami-$m$ fading channels for two and three users' scenarios, where
imperfect SIC is considered. The BER can be evaluated numerically for
general $m$ values, as one of the integrals does not have an analytical
solution. For the special case of Rayleigh fading, $m=1$, closed-form
expressions are derived for several cases of interest. Moreover, constrained
nonlinear optimization problems which aim to find the optimum power
coefficients that minimize the average BER and achieve fairness among the
users were formulated. Each objective can be used for different purposes,
for example, if the main goal of NOMA\ system is to fairly acquire equal
average BER\ for all users such as satellite systems, then the fairness
optimal power allocation should be evaluated. On the other hand, when the
purpose of a NOMA system is to minimize the overall average BER for
applications such as , then the latter problem formulation can be used.

\section*{\label{Sec-Appendix AVG BER}Appendix I: Average BER over Nakagami-$%
m$ Fading Channel}

The average BER for a NOMA\ system over Nakagami-$m$ fading channel follows
the order statistics of Nakagami-$m$ distribution. Based on order statistics
theory, the general ordered PDF of the channel gain of the $n$th user can be
expressed as \cite{7510798},%
\begin{equation}
f_{n}\left( \alpha_{n}\right) =K_{n}f\left( \alpha_{n}\right) \left[ F\left(
\alpha_{n}\right) \right] ^{n-1}\left[ 1-F\left( \alpha _{n}\right) \right]
^{N-n}  \label{generalOrder}
\end{equation}
where $K_{n}=\frac{N!}{\left( n-1\right) !\left( N-n\right) !}$, $f\left(
\alpha_{n}\right) $ and $F\left( \alpha_{n}\right) $ are respectively the
PDF\ and CDF of Nakagami-$m$ distribution with parameters $m$ and $\Omega$, 
\begin{equation}
f(\alpha_{n})=\frac{2m^{m}\alpha_{n}^{2m-1}}{\Omega^{m}\Gamma\left( m\right) 
}\mathrm{e}^{\left( -\frac{m\alpha_{n}^{2}}{\Omega}\right) }  \label{gggg}
\end{equation}%
\begin{equation}
F(\alpha_{n})=\frac{1}{\Gamma\left( m\right) }\Phi\left( m,\frac {%
m\alpha_{n}^{2}}{\Omega}\right)
\end{equation}
where $\Gamma(m)$ is the upper incomplete Gamma function, $\Omega =\mathbb{E}%
\left( \alpha_{n}^{2}\right) $, $m=\frac{\Omega^{2}}{Var\left(
\alpha_{n}^{2}\right) }$, and $\Phi\left( a,z\right) =\dint _{0}^{z}t^{a-1}%
\mathrm{e}^{-t}dt$ is the lower incomplete Gamma function \cite{Fraczek2017}%
. Therefore, the ordered PDF\ of the $n$th channel gain over Nakagami-$m$
channel is%
\begin{multline}
f_{n}(\alpha_{n})=\frac{2K_{n}m^{m}\alpha_{n}^{2m-1}}{\Omega^{m}\left[
\Gamma\left( m\right) \right] ^{n}}\mathrm{e}^{\left( -\frac{m\alpha _{n}^{2}%
}{\Omega}\right) }\left[ \Phi\left( m,\frac{m\alpha_{n}^{2}}{\Omega}\right) %
\right] ^{n-1}  \label{E-Ordered-PDF} \\
\times\left[ 1-\frac{\Phi\left( m,\frac{m\alpha_{n}^{2}}{\Omega}\right) }{%
\Gamma\left( m\right) }\right] ^{N-n}.
\end{multline}
Because the lower incomplete Gamma function is raised to a power, rendering
the integral analytically for $m>1$ is intractable. Therefore, the infinite
series representation of the lower incomplete Gamma function can be used 
\cite{6516881},%
\begin{equation}
\left[ \Phi\left( m,\frac{m\alpha_{n}^{2}}{\Omega}\right) \right] ^{\mu
}=\left( \frac{m\alpha_{n}^{2}}{\Omega}\right) ^{m\mu}\left[ \Gamma\left(
m\right) \right] ^{\mu}\mathrm{e}^{-\frac{\mu m\alpha_{n}^{2}}{\Omega}%
}\sum_{i=0}^{\infty}S_{i}\alpha_{n}^{2i}  \label{Incomple}
\end{equation}
where 
\begin{multline}
S_{i}=\left\{ 
\begin{array}{lc}
a_{0}^{\mu}\text{,} & i=0 \\ 
\frac{1}{ia_{0}}\dsum \limits_{z=1}^{i}\left( z\left( \mu+1\right) -i\right)
a_{z}S_{i-z}\text{, } & i\neq0%
\end{array}
\right. \text{, } \\
a_{_{z}}=\frac{\left( \frac{m}{\Omega}\right) ^{z}}{\Gamma(m+z+1)}\text{, }%
z=0,1,...,\infty.
\end{multline}
In addition, the term $\left[ 1-\frac{\Phi\left( m,\frac{m\alpha_{n}^{2}}{%
\Omega}\right) }{\Gamma\left( m\right) }\right] ^{N-n}$is expanded using
binomial theorem \cite{GRADSHTEYN198021} 
\begin{equation}
\left[ 1-\frac{\Phi\left( m,\frac{m\alpha_{n}^{2}}{\Omega}\right) }{%
\Gamma\left( m\right) }\right] ^{N-n}=\sum_{k=0}^{N-n}\binom{N-n}{k}\left( 
\frac{-\Phi\left( m,\frac{m\alpha_{n}^{2}}{\Omega}\right) }{\Gamma\left(
m\right) }\right) ^{k}.
\end{equation}
Now, $\gamma_{n,c}$ for a Nakagami-$m$ channel follows the Gamma
distribution $\mathcal{G}\left( k,\theta\right) $ where $k=m$, and $\theta=%
\frac{\Omega }{m}$, with the following PDF\ and CDF%
\begin{equation}
f(\gamma_{n,c})=\frac{m^{m}\gamma_{n,c}^{m-1}}{\overline{\gamma}%
_{n,c}^{m}\Gamma(m)}\mathrm{e}^{-\frac{m_{n}\gamma_{n,c}}{\bar{\gamma}_{n,c}}%
}  \label{ogammapdf}
\end{equation}
and 
\begin{equation}
F\left( \gamma_{n,c}\right) =\frac{\Phi\left( m,\frac{m\gamma_{n,c}}{%
\overline{\gamma}_{n,c}}\right) }{\Gamma(m)}
\end{equation}
respectively, where $\overline{\gamma}_{n,c}=A_{u_{2}u_{2}u_{3}}^{2}\Omega/%
\sigma_{n}^{2}$ and $c$ is the index parameter.

The ordered PDF\ of $\gamma_{n,c}$ of the $n$th channel can be expressed
using (\ref{generalOrder}), (\ref{Incomple}), and (\ref{ogammapdf}), as
follows%
\begin{multline}
f_{n}(\gamma_{n,c})=\frac{K_{n}}{\Gamma(m)}\sum_{k=0}^{N-n}\binom{N-n}{k}%
\left( -1\right) ^{k}\mathrm{e}^{-\frac{m\text{ }\gamma_{_{n,c}}}{\overline{%
\gamma}_{n,c}}\left( n+k\right) }  \label{gammaorder} \\
\times\left( \frac{m}{\overline{\gamma}_{n,c}}\right) ^{m(n+k)}\sum
_{i=0}^{\infty}S_{i}\gamma_{_{n,c}}^{i+m(n+k-1)}.
\end{multline}
In order to evaluate the average BER\ of the $n$th user over Nakagami-$m$
channel, (\ref{gammaorder}) and the alternative representation of the $Q$
function defined by \cite{258319} are utilized. Moreover, the following
integral is used \cite{GRADSHTEYN198021}, 
\begin{equation}
\int_{0}^{\infty}x^{t}\mathrm{e}^{-bx}dx=\frac{t!}{b^{t+1}}\text{, \ \ }%
t\in\left\{ 0,1,\ldots,b\right\} >0.  \label{tableofinte2}
\end{equation}
The general average BER for user $n$ is given by,%
\begin{align}
\overline{P}_{U_{n}} & =\int_{0}^{\infty}Q\left( \gamma_{n,c}\right)
f_{n}\left( \gamma_{n,c}\right) d\gamma_{n,c}  \notag \\
& =\frac{K_{n}}{\pi\Gamma(m)}\int\limits_{0}^{\infty}\int\limits_{0}^{\frac{%
\pi}{2}}\frac{1}{\mathrm{e}^{\frac{\gamma_{n,c}}{2\sin^{2}\left(
\psi_{n,c}\right) }}}\sum_{k=0}^{N-n}\tbinom{N-n}{k}\frac{\left( -1\right)
^{k}}{\mathrm{e}^{\frac{m\text{ }\gamma_{_{n},c}}{\overline{\gamma}_{n,c}}%
\left( n+k\right) }}  \notag \\
& \text{ \ \ \ \ \ \ \ }\times\left( \frac{m}{\overline{\gamma}_{n,c}}%
\right)
^{m(n+k)}\sum_{i=0}^{\infty}S_{i}\gamma_{_{n,c}}^{i+m(n+k-1)}d\psi_{n,c}d%
\gamma_{n,c}  \notag \\
& =\frac{K_{n}}{\pi\Gamma(m)}\sum_{k=0}^{N-n}\sum_{i=0}^{\infty}\tbinom {N-n%
}{k}\left( -1\right) ^{k}S_{i}\frac{m^{m(n+k)}}{\overline{\gamma}%
_{n,c}^{m(n+k)}}  \notag \\
& \text{ \ \ \ \ \ \ }\times\int\limits_{0}^{\frac{\pi}{2}}\frac{\left[
i+m(n+k-1)\right] !}{\left( \frac{1}{2\sin^{2}\left( \psi_{n,c}\right) }+%
\frac{m\text{ }\left( n+k\right) }{\overline{\gamma}_{n,c}}\right)
^{i+m(n+k-1)+1}}d\psi_{n,c}.  \label{E-PU1_N2}
\end{align}
Although the integral in (\ref{E-PU1_N2}) does not have an analytical
solution, it can be easily solved numerically.

\bibliographystyle{IEEEtran}
\bibliography{myref}

\textbf{Tasneem Assaf} (S'14) received the B.Sc. degree in communications
engineering from Khalifa University (KU), Abu Dhabi, UAE, in 2014, and the
master's degree in electrical engineering from the American University of
Sharjah (AUS), Sharjah, UAE, in 2016. She is currently pursuing the Ph.D.
degree from KU. Her research interests include smart grids, optimization,
wireless communications, and machine learning.

\textbf{Arafat Al-Dweik} (S'97-M'01-SM'04) received the M.S. (Summa Cum
Laude) and Ph.D. (Magna Cum Laude) degrees in electrical engineering from
Cleveland State University, Cleveland, OH, USA in 1998 and 2001,
respectively. He was with Efficient Channel Coding, Inc., Cleveland-Ohio,
from 1999 to 2001. From 2001 to 2003, he was the Head of Department of
Information Technology at the Arab American University in Palestine. Since
2003, he is with the Department of Electrical Engineering, Khalifa
University, United Arab Emirates. He joined University of Guelph, ON,
Canada, as an Associate Professor from 2013-2014. Dr. Al-Dweik is a Visiting
Research Fellow at the School of Electrical, Electronic and Computer
Engineering, Newcastle University, Newcastle upon Tyne, UK and he is
Research Professor Western University, London, ON, Canada and University of
Guelph. Dr Al-Dweik has extensive editorial experience where he served as an
Associate Editor at the IEEE Transactions on Vehicular Technology and the
IET Communications. Dr Al-Dweik has extensive research experience in various
areas of wireless communications that include modulation techniques, channel
modeling and characterization, synchronization and channel estimation
techniques, OFDM technology, error detection and correction techniques, MIMO
and resource allocation for wireless networks. Dr. Al-Dweik is a member of
Tau Beta Pi and Eta Kappa Nu. He was awarded the Fulbright scholarship from
1997 to 1999. He received the Hijjawi Award for Applied Sciences in 2003 and
the Fulbright Alumni Development Grant in 2003 and 2005, and the Dubai Award
for Sustainable Transportation in 2016. He is a Senior Member of the IEEE,
and a Registered Professional Engineer in the Province of Ontario, Canada.

\textbf{Mohamed El Moursi} (M'12, SM15) received the B.Sc. and M.Sc. degrees
from Mansoura University, Mansoura, Egypt, in 1997 and 2002, respectively,
and the Ph.D. degree from the University of New Brunswick (UNB),
Fredericton, NB, Canada, in 2005, all in electrical engineering. He was a
Research and Teaching Assistant in the Department of Electrical and Computer
Engineering, UNB, from 2002 to 2005. He joined McGill University as a
Postdoctoral Fellow with the Power Electronics Group. He joined Vestas Wind
Systems, Arhus, Denmark, in the Technology R\&D with the Wind Power Plant
Group. He was with TRANSCO, UAE, as a Senior Study and Planning Engineer. He
is currently a Professor in the Electrical and Computer Engineering
Department at Khalifa University of Science and Technology- Masdar Campus
and seconded to a Professor Position in the Faculty of Engineering, Mansoura
University, Mansoura, Egypt and currently on leave. He was a Visiting
Professor at Massachusetts Institute of Technology, Cambridge,
Massachusetts, USA. Dr. Shawky is currently an Editor of IEEE Transactions
on Power Delivery, IEEE Transactions on Power Systems, Associate Editor of
IEEE Transactions on Power Electronics, Guest Editor of IEEE Transactions on
Energy Conversion, Guest Editor-in-Chief for special section between TPWRD
and RPWRS, Editor for IEEE Power Engineering Letters, Regional Editor for
IET Renewable Power Generation and Associate Editor for IET Power
Electronics Journals. His research interests include power system, power
electronics, FACTS technologies, VSC-HVDC systems, Microgrid operation and
control, Renewable energy systems (Wind and PV) integration and
interconnections.

\textbf{Hatem Zeineldin} (M'06--SM'13) received the B.Sc. and M.Sc. degrees
in electrical engineering from Cairo University, Giza, Egypt, in 1999 and
2002, respectively, and the Ph.D. degree in electrical and computer
engineering from the University of Waterloo, Waterloo, ON, Canada, in 2006.
He was with Smith and Andersen Electrical Engineering, Inc., North York, ON,
USA, where he was involved in projects involving distribution system
designs, protection, and distributed generation. He was a Visiting Professor
with the Massachusetts Institute of Technology, Cambridge, MA, USA. He is
currently a Professor with the Khalifa University of Science and Technology,
Abu Dhabi, UAE, and is currently on leave from the Faculty of Engineering,
Cairo University, Giza. His current research interests include distribution
system protection, distributed generation, and micro grids. He is currently
an Editor for the IEEE Transactions on Energy Conversion and the IEEE
Transactions on Smart Grid.

\end{document}